\documentclass[reprint,aps,prl,amsmath,amssymb,superscriptaddress]{revtex4-2} 
\usepackage{graphicx} 
\usepackage{bm} 
\usepackage[version=4]{mhchem}
\usepackage[hidelinks]{hyperref} 
\usepackage{physics} 
\begin{document} 
\title{Perfect Absorption in the Strong Coupling Regime via Degenerate Critical Coupling}
\author{Eleonora P. Kraus} 
\email{eleonora.kraus@physik.uni-marburg.de} 
\affiliation{Department of Physics, Philipps-Universit\"{a}t Marburg, 35037 Marburg, Germany} 
\affiliation{mar.quest|Marburg Center for Quantum Materials and Sustainable Technologies, 35032 Marburg, Germany} 
\author{Jamie M. Fitzgerald} 
\affiliation{Department of Physics, Philipps-Universit\"{a}t Marburg, 35037 Marburg, Germany} 
\affiliation{mar.quest|Marburg Center for Quantum Materials and Sustainable Technologies, 35032 Marburg, Germany} 
\author{Carlos Maciel-Escudero} 
\affiliation{Department of Physics, Philipps-Universit\"{a}t Marburg, 35037 Marburg, Germany} 
\affiliation{mar.quest|Marburg Center for Quantum Materials and Sustainable Technologies, 35032 Marburg, Germany} 
\author{Ermin Malic} 
\affiliation{Department of Physics, Philipps-Universit\"{a}t Marburg, 35037 Marburg, Germany} 
\affiliation{mar.quest|Marburg Center for Quantum Materials and Sustainable Technologies, 35032 Marburg, Germany}

\begin{abstract}
    Perfect absorption (PA) represents a fundamental limit of light-matter interaction and a means to maximize nanoscale energy conversion. While PA is now a well-established phenomenon, both the theoretical feasibility and a practical mechanism for achieving it under single-beam excitation within the strong coupling regime is unknown. Through rigorous solution of Maxwell’s equations for a compact photonic crystal (PhC) architecture incorporating a two-dimensional semiconductor, we present a general method based on degenerate critical coupling for single-port PA of exciton-polaritons. At the crossing of two polariton branches, we achieve near-unity absorption exceeding 99.8 \% in a structure thinner than $100\,$nm. This effect is robust under realistic Gaussian beam excitation, and can be realized across different temperatures and excitonic materials by tailoring the PhC geometry. Our results establish a strategy for enabling efficient light–matter coupling, with direct implications for the development of metal-free, ultra-compact polaritonic logic devices, sensors, and energy-harvesting platforms. 
\end{abstract}

\maketitle
\newpage
\section{Introduction} 

The efficient conversion of electromagnetic energy into electronic excitations is essential for both characterizing material systems and exploiting light-matter interactions for optoelectronic applications. In highly compact devices, where conventional Beer-Lambert attenuation is insufficient, coherent wave interference provides a powerful mechanism to maximize energy transfer \cite{kats2016optical,baranov2017coherent, li2019engineering}. The ultimate realization of this principle is perfect absorption (PA): a phenomenon where a system with loss is precisely engineered so that all outgoing waves destructively interfere, completely trapping the incident radiation to achieve 100\% absorption \cite{chong2010coherent,wan2011time,noh2012perfect,zhang2012controlling}. As PA relies on far-field and/or modal interference rather than material-specific resonances, this allows, with suitable geometric engineering, for efficient energy trapping in ultra-thin and otherwise weakly-absorbing media, such as two-dimensional (2D) semiconductors \cite{horng2019engineering,epstein2020near,horng2020perfect,lee2023achieving,canales2023perfect} and graphene \cite{thongrattanasiri2012complete,zhang2014coherent,piper2014total,piper2014total_2,rao2014coherent,thareja2015electrically}. Beyond its fundamental significance and direct connection to lasing \cite{chong2010coherent,wan2011time} and thermal emission \cite{baranov2019nanophotonic}, perfect absorption is highly relevant for photovoltaics \cite{polman2012photonic,liew2016coherent}, and photodetection \cite{kishino1991resonant,jin2021organic}. Recently, PA has been extended to arbitrary wavefronts \cite{slobodkin2022massively} and time-varying media \cite{galiffi2026optical}. Furthermore, it is applicable over a wide range of wavelengths in optics, from microwave \cite{li2014equivalent} to visible \cite{piper2014total_2}, as well as to more general wave phenomena, such as sound \cite{jimenez2016ultra} and water waves \cite{euve2023perfect}. 

\begin{figure}[t!]
    \centering
    \includegraphics[width=\linewidth]{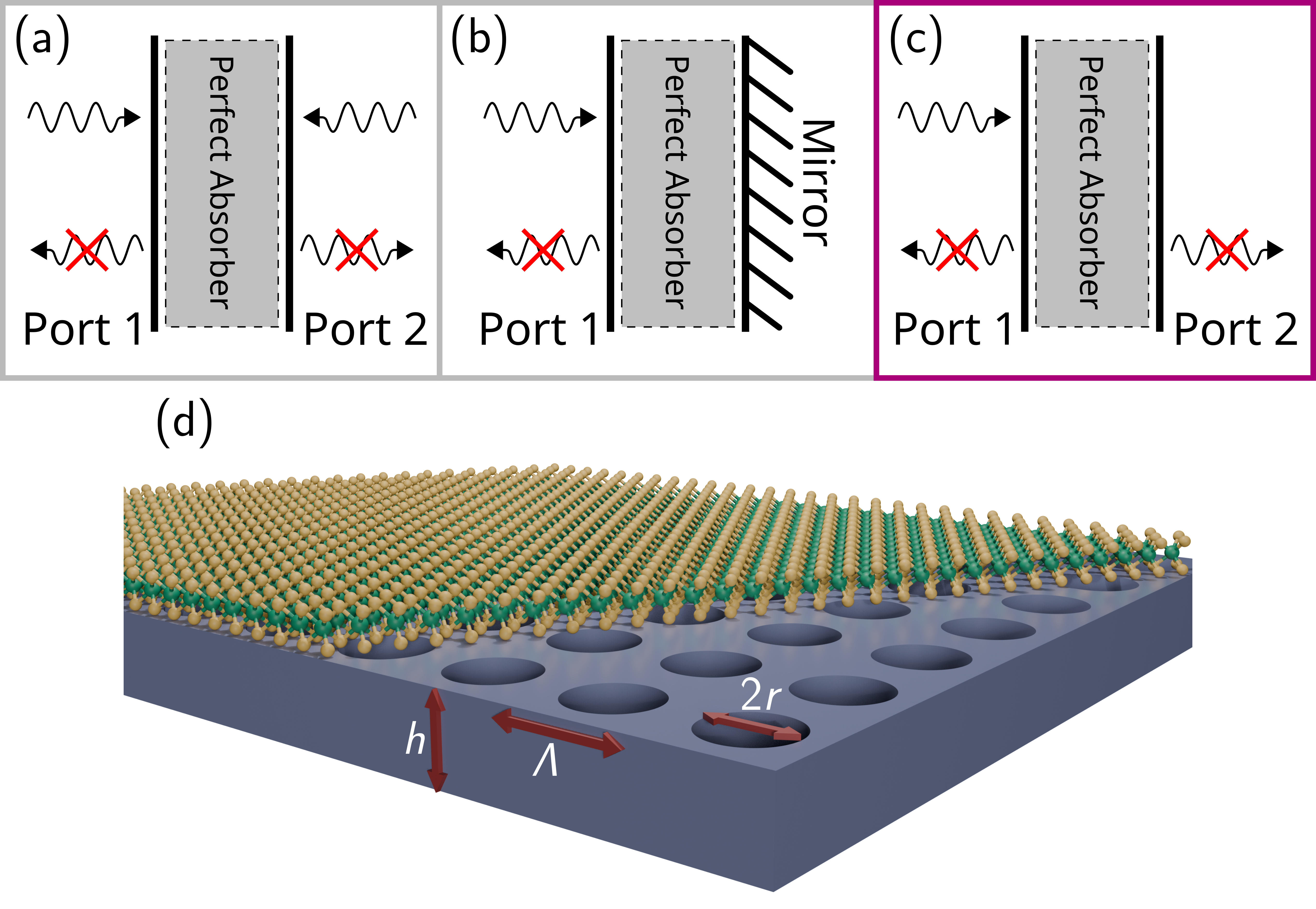}
    \caption{Schematic depiction of three configurations for achieving total absorption: \textbf{(a)} two-port system excited by two beams, \textbf{(b)} one-port system excited by one beam, and \textbf{(c)}  two-port system excited by one beam, which corresponds to the system explored in this work. Note that each outgoing crossed arrow means that no light leaves the system through the ports (perfect absorber). \textbf{(d)} Schematic of a \ce{WS2} monolayer on a silicon photonic crystal (PhC) slab patterned with a square lattice of air holes. }
    \label{fig:Schematics}
\end{figure}

Experimental realizations of PA have typically required either two-port configurations that demand precise interferometric alignment between two counter-propagating waves (coherent perfect absorption, Fig.~\ref{fig:Schematics}a) \cite{chong2010coherent,wan2011time,zhang2014coherent} or bulky one-port geometries with a highly reflective back mirror and a single incident channel (a Salisbury screen, Fig.~\ref{fig:Schematics}b) \cite{piper2014total_2, thareja2015electrically,horng2019engineering,epstein2020near,horng2020perfect,lee2023achieving}, limiting their utility for many practical applications. 
Structured dielectric media, such as photonic crystal (PhC) slabs \cite{piper2014total_2,piper2016broadband,zhou2016perfect} and metasurfaces \cite{watts2012metamaterial}, have emerged as promising platforms to achieve subwavelength PA with single-port excitation, offering a path towards high-efficiency optoelectronic devices that are compact, metal free, and highly tuneable. In particular, PhCs that support sharp, asymmetric Fano resonances \cite{miroshnichenko2010fano} can achieve narrow-band PA, suitable for applications in ultra-sensitive sensing and optical switches. For example, Piper \emph{et al} proposed the mechanism of degenerate critical coupling to achieve 100\% absorption for a graphene monolayer on top of a mirror-symmetric PhC slab \cite{piper2014total}, where the concept has since been experimentally verified \cite{liu2017experimental}.

A primary objective in nanophotonics is the efficient transfer of electromagnetic energy into active semiconductor monolayers \cite{jariwala2016near,li2019engineering}. Two-dimensional semiconductors, such as transition metal dichalcogenides (TMDs), support room-temperature-stable excitons with exceptionally large oscillator strengths \cite{wang2018colloquium,perea2022exciton,mueller18}. By integrating these materials with PhCs, the spatial confinement of the optical fields can be leveraged to drive light-matter interactions into the strong-coupling regime \cite{zhang2018photonic,kravtsov2020nonlinear,chen2020metasurface,he2023polaritonic,maggiolini2023strongly, kraus2026engineering}. This interaction gives rise to exciton-polaritons—hybrid light-matter quasi-particles that combine the long-range coherence of photons with the strong nonlinearity of excitons \cite{fitzgerald2025polariton}, providing a promising route to compact, ultra-fast photonic circuitry \cite{genco2025femtosecond}. Despite the rich variety of photonic systems where PA has been demonstrated, it remains largely unexplored in the strong-coupling regime. In a pioneering study \cite{zanotto2014perfect}, near-perfect polaritonic absorption under two-port excitation was demonstrated by engineering the critical coupling condition for a \ce{AlGaAs/GaAs} quantum well heterostructure patterned with periodic gold stripes. Furthermore, patterned slabs of bulk \ce{WS2} have been predicted to support PA of self-hybridized exciton-polaritons under two-port excitation \cite{gu2023polaritonic}. However, demonstration of one-port PA in the strong coupling regime remains elusive. Extending PA to the strong-coupling regime offers potential for the efficient generation of exciton-polaritons for lasing and ultra-low power optical switches, as well as  enabling electric \cite{mondal2025switching} and magnetic \cite{konig2025magneto} field control. 

In this theoretical work, we solve the full Maxwell’s equations with realistic material parameters to investigate PA in the strong-coupling regime for a \ce{WS2} monolayer integrated on top of a PhC slab. By exploiting the parity of the PhC modes, we provide, to the best of our knowledge, the first demonstration of PA via degenerate critical coupling within the strong-coupling regime for a metal-free structure under single-port excitation. Using coupled-mode theory, we show that the system operates in the ``photon-decoupled regime'', characterized by two pairs of independent polariton branches. Lastly, we demonstrate that PA is achievable over a wide range of temperatures by engineering the PhC geometry. This works thus provides an important step towards designing polaritonic devices that exploit PA to achieve highly efficient in-coupling of energy.

%%%%%%%%%%%%%%
% Section 1: %
%%%%%%%%%%%%%%

\section{Perfect Absorption via Degenerate Critical Coupling}

\begin{figure}[t!]
    \centering
    \includegraphics[width=\linewidth]{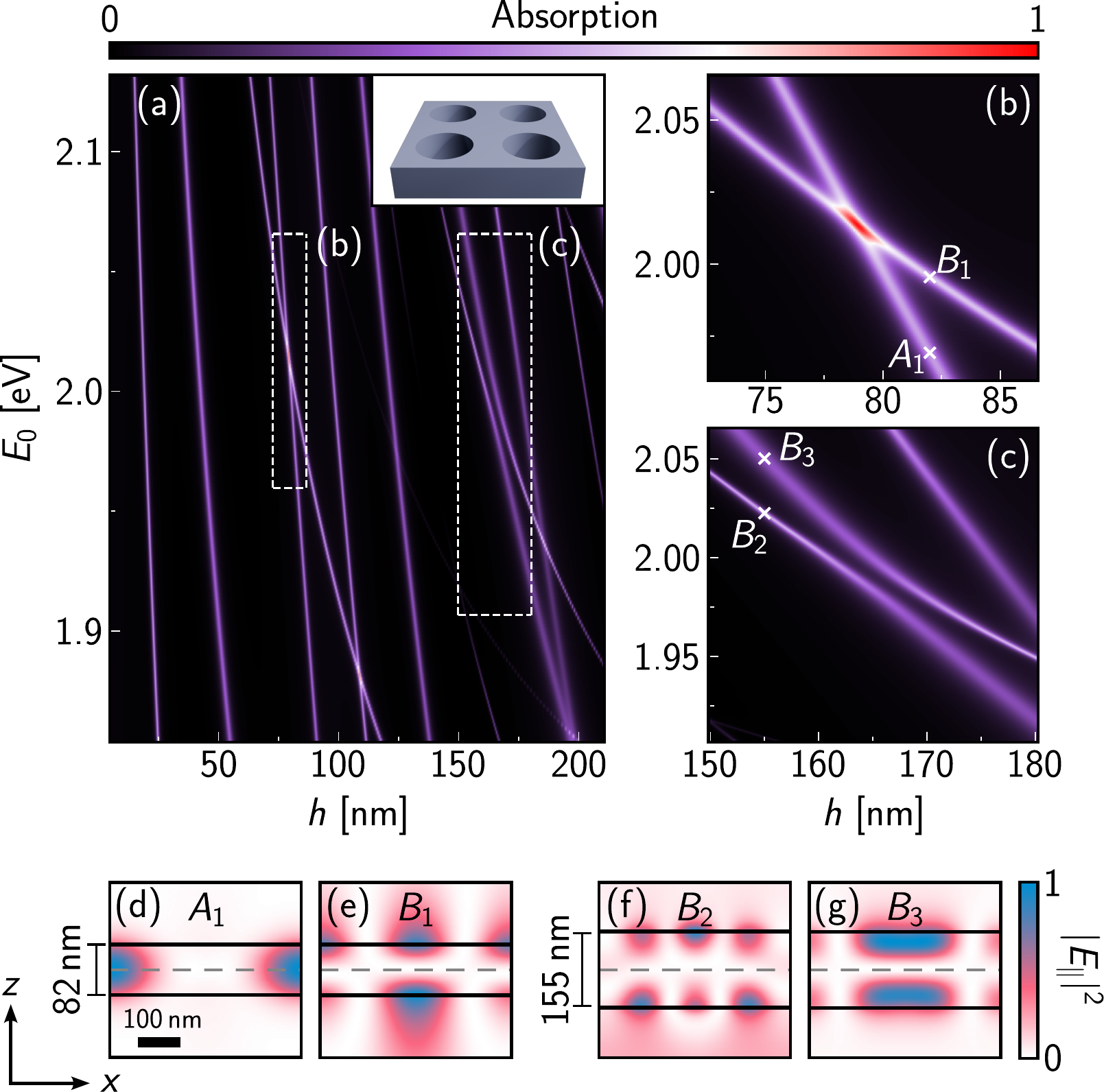}
    \caption{\textbf{(a)} Absorption spectra calculated with RCWA over incident energy $E_0$ and PhC thickness $h$ at normal incidence for fixed $\Lambda=468\,$nm and $r=89\,$nm, corresponding to the filling factor $f=1 - \pi r^2 / \Lambda^2 = 0.88$. Two different behaviors of intersecting photonic modes can be observed: Crossings and avoided crossings. One of each highlighted by the white dashed boxes. \textbf{(b)} Closeup of the crossing of two photonic modes labeled A$_1$ and B$_1$ as highlighted in (a). A maximum absorption of $A\approx100\,\%$ is observed. \textbf{(c)} Closeup of the anti-crossing of two photonic modes labeled B$_2$ and B$_3$ as highlighted in (a). Here the maximum absorption is only $A\approx50\,\%$. \textbf{(d) - (g)} Normalized in-plane electric field amplitude of the four modes A$_1$ and B$_{1,2,3}$ indicated by white crosses in (b) and (c) shown in a $xz$-slice at the edge of the unit cell. Each mode is classified as either even (A$_n$) or odd (B$_n$) with respect to the plane of symmetry of the PhC (gray dashed line at the center of the PhC slab). The scale bar in the bottom left corner of panel (d) applies to all panels.} 
    \label{fig:Mode_Characterization}
\end{figure}

In this work, we focus on a sub-wavelength thick silicon-based 2D PhC slab ($\varepsilon_\text{\ce{Si}}=12.39$) patterned with a periodic array of air holes, as shown in Fig.~\ref{fig:Schematics}d. The hole radius $r$, slab thickness $h$, and lattice period $\Lambda$ constitute three independent geometrical parameters to tune the optical response of the PhC. All numerical simulations presented below were performed using rigorous coupled-wave analysis (RCWA), a semi-analytical method ideal for efficiently modeling layered structures with in-plane periodicity \cite{whittaker1999scattering}. 
First, to characterize the PhC modes and the degenerate critical coupling condition in the absence of an exciton response, we introduce an artificial loss ($\Im{\varepsilon_\text{\ce{Si}}} = -3$) to the dielectric function of the bare PhC slab. This introduces a material-based loss, resulting in Lorentzian lineshapes in absorption that provide a clearer visualization of the modes than the Fano-like profiles observed in reflection and transmission spectra. Figure \ref{fig:Mode_Characterization}a shows the resulting absorption spectra as a function of incident energy $E_0$ and slab thickness $h$. The distinct bands in the spectra correspond to the excitation of the photonic eigenmodes of the 2D PhC, which exhibit a strong dependence on the thickness. 

Owing to the mirror symmetry of the 2D PhC, these eigenmodes can be classified into two distinct, orthogonal sets with well-defined parity, denoted as A$_n$ and B$_n$ for even and odd modes, respectively, where the label $n$ indicates energetic ordering. Modes of different parity do not couple, resulting in a crossing of their dispersive branches, as exemplified for the A$_1$ and B$_1$ modes in Fig.~\ref{fig:Mode_Characterization}b. In contrast, modes with the same parity can hybridize, leading to avoided crossings of their dispersive branches, as illustrated in Fig.~\ref{fig:Mode_Characterization}c. The parity of the photonic modes can be directly identified from the symmetry of the electric field profile with respect to the PhC mid-plane (gray dashed line in Figs.~\ref{fig:Mode_Characterization}d-g). Here, the normalized in-plane electric field amplitude profiles are shown in the $xz$-plane at the edge of the PhC unit cell, evaluated at the points indicated by the white crosses in Figs.~\ref{fig:Mode_Characterization}b-c. The A$_1$ mode exhibits a strong confinement of the electric field within the PhC slab, whereas the B modes are more delocalized. All modes show spatial variation within the in-plane direction. The parity of the PhC modes is determined by the polarization (transverse electric or magnetic) of the underlying fundamental slab waveguide modes, which are zone-folded into the lightcone by the in-plane periodicity of the PhC slab, see SI Note 1.

The mirror symmetry of the PhC slab determines the maximum absorption. Symmetric two-port systems, such as the one shown in Fig.~\ref{fig:Mode_Characterization} and schematically in Fig.~\ref{fig:Schematics}c, exhibit an absorption limit of $50\,\%$ \emph{per mode} when excited from a single port, i.e., from one side of the slab. This can be understood from the parity of the photonic eigenmodes: the incident illumination can be decomposed into equal contributions of even and odd symmetry, of which only the component with matching parity can couple to a given mode \cite{piper2014total}. Total absorption under single-port excitation is possible if the system supports two energy-degenerate modes of opposite parity. When both modes independently satisfy the critical coupling condition, where internal dissipation exactly balances the radiative coupling rate, their combined interference simultaneously suppresses both reflection and transmission. The system considered in Fig.~\ref{fig:Mode_Characterization}a can reach an absorption of approximately $100\,\%$ since two modes of opposite parity become energetically degenerate at a slab thickness of $h=78.9\,$nm and simultaneously satisfy the critical coupling condition, as shown in Fig.~\ref{fig:Mode_Characterization}b. This is known as degenerate critical coupling \cite{piper2014total}. %In the next section, we investigate whether PA via degenerate critical coupling can also be achieved in the strong-coupling regime, enabling efficient excitation of polaritonic states.

%%%%%%%%%%%%%%
% Section 2: %
%%%%%%%%%%%%%%

\section{Realizing Perfect Absorption with Exciton-Polaritons}
\begin{figure}[t!]
    \centering
    \includegraphics[width=\linewidth]{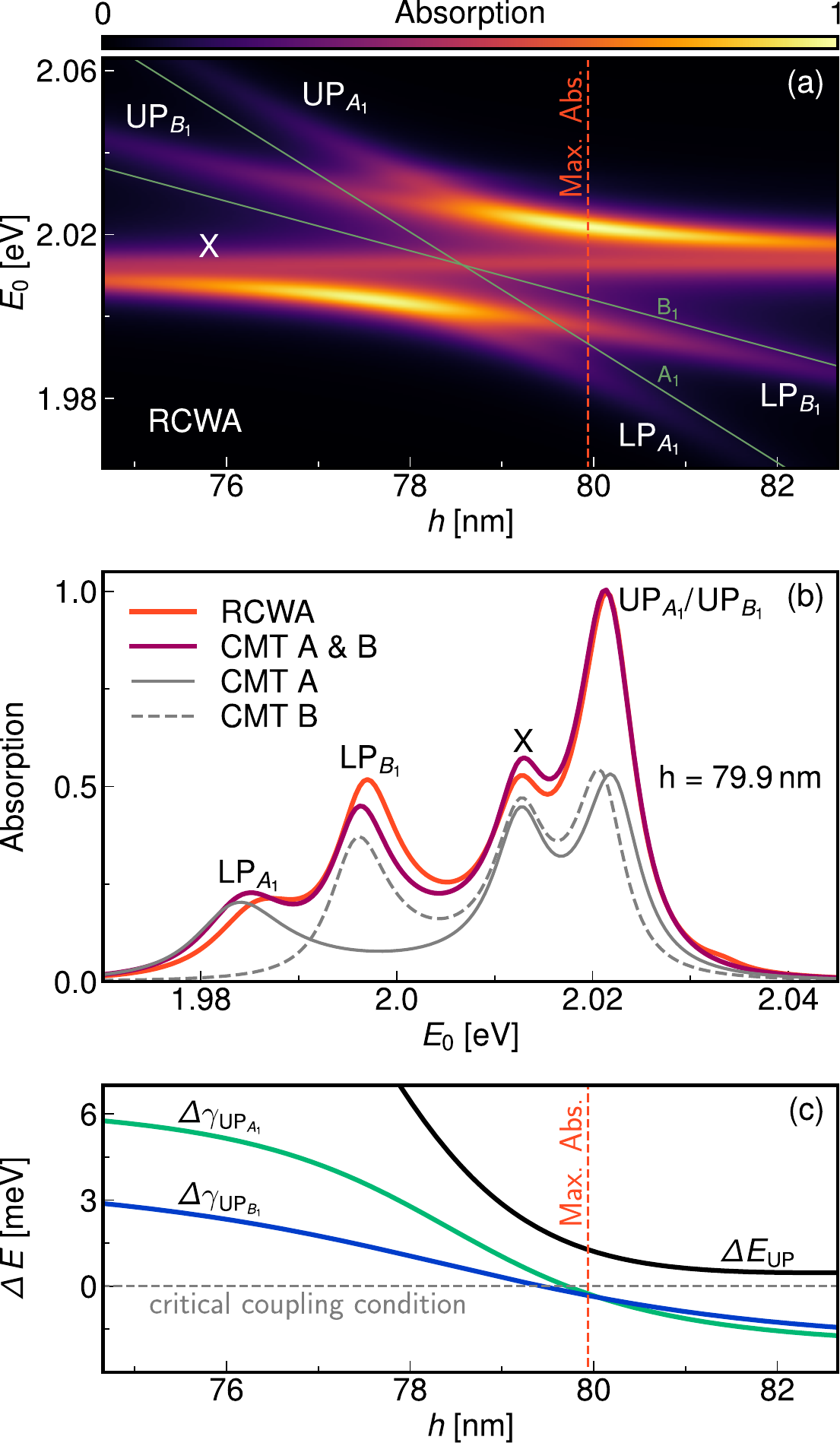}
    \caption{\textbf{(a)} Absorption spectra calculated with RCWA at normal incidence for the PhC slab with a \ce{WS2} monolayer on top. The spectra are plotted as a function of incident energy $E_0$ and PhC thickness $h$. Five distinct modes can be observed: the weakly coupled exciton $X$ and two sets of upper (UP$_{\text{A}_1,\text{B}_1}$) and lower (LP$_{\text{A}_1,\text{B}_1}$) polariton branches. The subscript A$_1$/B$_1$ refers to the polaritons that form through strong coupling of the exciton to the photonic mode A$_1$/B$_1$, as indicated by the green lines. \textbf{(b)} Absorption spectrum of the system at $h = 79.9$\,nm (corresponding to the red dashed line in panel (a)) comparing the RCWA (orange line) with the CMT calculation with both photonic modes included (purple line) and only one photonic mode included (gray lines). \textbf{(c) }
    \label{fig:RCWA_CMT_Comparison} Difference of the spectral position of the UP branches $\Delta E_\text{UP} = E_{\text{UP}_{\text{A}_1}} - E_{\text{UP}_{\text{B}_1}}$ (black line) and of the UP decay rates $\Delta \gamma_{\text{UP}_{\text{A}_1/\text{B}_1}} = \Gamma_{\text{UP}_{\text{A}_1/\text{B}_1}} - \gamma_{\text{UP}_{\text{A}_1/\text{B}_1}}$ (green/blue line) over the PhC thickness $h$. For  $\Delta \gamma_{\text{UP}_{\text{A}_1/\text{B}_1}} = 0$, UP$_{\text{A}_1/\text{B}_1}$ is critically coupled as indicated by the gray dashed line.}
\end{figure}

To investigate the possibility of PA in the strong-coupling regime, we introduce an excitonic TMD monolayer atop the PhC slab. We choose \ce{WS2} as a representative example due to its large excitonic oscillator strength. However, our results remain generalizable to other excitonic materials. The optical response of \ce{WS2} is characterized by the dielectric function $\varepsilon_{\ce{WS2}}$ (see Methods), which captures the excitonic resonances in the visible range. Due to the monolayer's negligible thickness ($h_{\text{\ce{WS2}}} \approx 0.6$ nm), its addition on top of the PhC slab approximately preserves the mirror symmetry of the structure, thereby maintaining the orthogonality of the even and odd photonic eigenmodes \cite{piper2014total}.
Figure~\ref{fig:RCWA_CMT_Comparison}a shows the absorption spectra of the system as a function of the incident energy $E_0$ and slab thickness $h$. Four branches emerge, corresponding to two upper polaritons (UP$_{\text{A}_1}$, UP$_{\text{B}_1}$) and two lower polaritons (LP$_{\text{A}_1}$, LP$_{\text{B}_1}$) resulting from the strong coupling between the two photonic modes A$_1$ and B$_1$ and the exciton state $X$ in the monolayer. The latter appears as a bright, flat branch at the energy of $2.013\,$eV, corresponding to an exciton resonance weakly coupled to the photonic modes \cite{kraus2026engineering}. Importantly, a maximum absorption of $99.8\,\%$ is achieved at $h \approx 79.9\,$nm, where the two upper polariton branches become energetically degenerate. At a reduced thickness ($h \approx 77.5\,$nm), corresponding to the crossing point of the two lower polariton branches, the system also exhibits near-unity absorption, reaching a maximum value of $A = 97.8\,\%$. These two results demonstrate that PA can be achieved within the strong-coupling regime. We note that the silicon PhC is assumed to be lossless, thus the absorption is localized entirely within the TMD monolayer, despite being less than 1 nm in thickness. Moreover, the relevant photonic modes exhibit strong spatial variation of the electric field within the $xy$ plane, creating "hot-spots" of localized absorption on a 100 nm length scale (see SI Note 4).

To gain a deeper understanding of the physical mechanism underlying the PA shown in Fig.~\ref{fig:RCWA_CMT_Comparison}a, we employ temporal coupled-mode theory (CMT) \cite{fan2003temporal} to reproduce the absorption spectra, see Methods and SI Note 2. When modeling the coupling between the exciton and the two relevant orthogonal PhC modes within CMT, it is important to set up the coupling between the modes correctly. Multi-mode systems with $N$ photonic modes and one excitonic resonance can be split up into two distinct cases \cite{balasubrahmaniyam2021coupling}: (i) the photon-coupled regime, where optical modes become indirectly coupled with each other by collectively coupling to the same excitonic resonance, giving rise to $N+1$ polariton branches; and (ii) the photon-decoupled regime, where each photonic mode couples independently to separate excitonic states. The latter is characterized by $2N$ polariton branches and a polariton bandgap, i.e., polariton branches do not cross the bare exciton resonance. This is the case in Fig.~\ref{fig:RCWA_CMT_Comparison}a, confirming that our system is described by a photon-decoupled Hopfield matrix, where the two photonic modes and their corresponding pair of polariton branches remain independent. This decoupled behavior can be understood as emerging from distinct dark slab waveguide polaritons of differing momenta that are zone-folded into the lightcone by the PhC periodicity, see SI Note 1. This behavior has been observed in microcavities filled with organic semiconductors characterized by large oscillator strengths \cite{georgiou2021observation}. Here, we demonstrate that it also arises in a fundamentally different material platform, driven not by the excitonic oscillator strength, but rather by the periodicity of the PhC slab.

The signature at the bare exciton energy in Fig.~\ref{fig:RCWA_CMT_Comparison}a (flat band at $2.013\,$eV) arises from strong spatial variations of the electric field distribution within the unit cell, leading to coexisting regions of weak and strong coupling between the excitonic resonance and the photonic modes on subwavelength scales \cite{kraus2026engineering}. From CMT, we can extract the decay rates, coupling strengths $g_{\text{A}_1/\text{B}_1}$ between the exciton and the two photonic modes, as well as the Hopfield coefficients $C_\text{X,Ph}$, which quantify the contribution of the constituent exciton and photons in the exciton-polaritons.
Figure~\ref{fig:RCWA_CMT_Comparison}b shows the comparison between the CMT (purple line) and the RCWA simulation (orange line) for $h = 79.9\,$nm, where a good agreement between the two can be observed. For this PhC thickness, the coupling strengths extracted from the CMT are $g_{\text{A}_1} = 16.86$\,meV and $g_{\text{B}_1} = 12.04$\,meV. This confirms that both photonic modes are strongly coupled to the excitonic resonance, with each individually satisfying the strong-coupling condition $g>\frac{\gamma_\text{Ph} + \gamma_{\text{X}}}{4}$ \cite{schneider2018two}.

To further elucidate the mechanism of PA via degenerate critical coupling, we use CMT to compute the individual polariton absorption spectra resulting from photonic modes $\text{A}_1$ and $\text{B}_1$ (gray lines). The maximum absorption is limited to $53.3\,\%$, consistent with the absorption of a single mode in a symmetric two-port system excited through one port, where the residual $3.3\,\%$ absorption is contributed by nearby resonances. This result demonstrates that achieving PA requires the simultaneous excitation of two energetically degenerate polariton branches. Further insight can be provided by analyzing the polariton decay rates, as PA requires that both polaritons individually satisfy the critical coupling condition $\gamma_{\text{P}_i} = \Gamma_{\text{P}_i}$, i.e., the polariton's radiative decay, $\gamma_{\text{P}_i}$, is balanced by exciton-based loss, $\Gamma_{\text{P}_i}$. The polariton decay rates are given by
\begin{equation}
    \gamma_{\text{P}_i} = |C_\text{X}|^2\gamma_{\text{X}} + |C_\text{Ph}|^2\gamma_{\text{Ph}_i}\,,\quad \Gamma_{\text{P}_i} = |C_\text{X}|^2\Gamma_{\text{X}}\,,\label{eq:polariton_decay}
\end{equation}
where $\Gamma_{\text{X}}$ is the non-radiative decay of the excitonic states, and $\gamma_{\text{P/X}}$ denotes the radiative decay rates of the photonic/excitonic modes. Note that the photonic modes do not contribute to the non-radiative decay rate of the polariton as the small material loss of the silicon is neglected (Im$(\varepsilon_{\ce{Si}})=0.02$) \cite{aspnes1983dielectric}. To demonstrate that the critical coupling condition is the relevant mechanism for PA, we show the difference in decay rates $\Delta \gamma_\text{P} = \Gamma_{\text{P}} - \gamma_{\text{P}}$ as a function of the PhC thickness in Fig.~\ref{fig:RCWA_CMT_Comparison}c. The green/blue solid line represents the decay rate difference of the upper A$_1$/B$_1$ polariton branch, respectively, which crosses zero when the critical coupling condition is satisfied. Additionally, to achieve \emph{degenerate} critical coupling, the two upper polaritons must have closely matched energies, which is depicted by their difference $\Delta E = E_{\text{A}_1} -E_{\text{B}_1}$ (black line). Maximum absorption is observed when the system best satisfies the simultaneous conditions of energetic degeneracy $(\Delta E_\text{UP} = 0)$ and individual critical coupling $(\Delta \gamma_\text{UP} = 0)$, as captured by the following polariton Elliott formula \cite{fitzgerald2022twist}:
\begin{equation}
A(\omega) = \sum_{i=\text{A,B},P=\text{A,B}}\frac{2 \gamma_{P,i} \Gamma_{P,i}}{\left(\gamma_{P,i} + \Gamma_{P,i}\right)^2 + (\hbar\omega - E_i)^2}.\label{eq:Elliot}
\end{equation}
The equation is derived using CMT under the approximation that the absorption is determined only by two sets of independent polaritons (see SI Note 2). The absorption $A(\omega)$ reaches unity when both modes are critically coupled and energetically degenerate.
The two photonic modes generally exhibit different linewidths ($\gamma_\text{A} \neq \gamma_\text{B}$), therefore one would expect that simultaneous critical coupling of both polaritons is nontrivial. However, this imbalance can be compensated by differences in the Hopfield coefficients of the polariton branches, provided that $\gamma_\text{Ph} \sim \gamma_\text{X}$, see Eq.~(\ref{eq:polariton_decay}). 
Note that the coupling strength between the photonic modes and the exciton resonance can be a tuning parameter to control the spectral position and the necessary PhC thickness for PA. By increasing the coupling strength between either of the photonic modes and the exciton, the Rabi splitting between the polariton branches increases and the upper/lower polariton shifts to higher/lower energies. Consequently, the spectral position of PA is modified as the crossing point of the two upper/lower polariton branches shifts. In practice, this can be achieved by engineering the PhC slab to maximize the light-matter coupling with the active semiconductor.
\cite{kraus2026engineering}.

%%%%%%%%%%%%%%
% Section 3: %
%%%%%%%%%%%%%%

\section{Temperature Dependence and Tunability of Perfect Absorption}
\begin{figure}[t!]
    \centering
    \includegraphics[width=\linewidth]{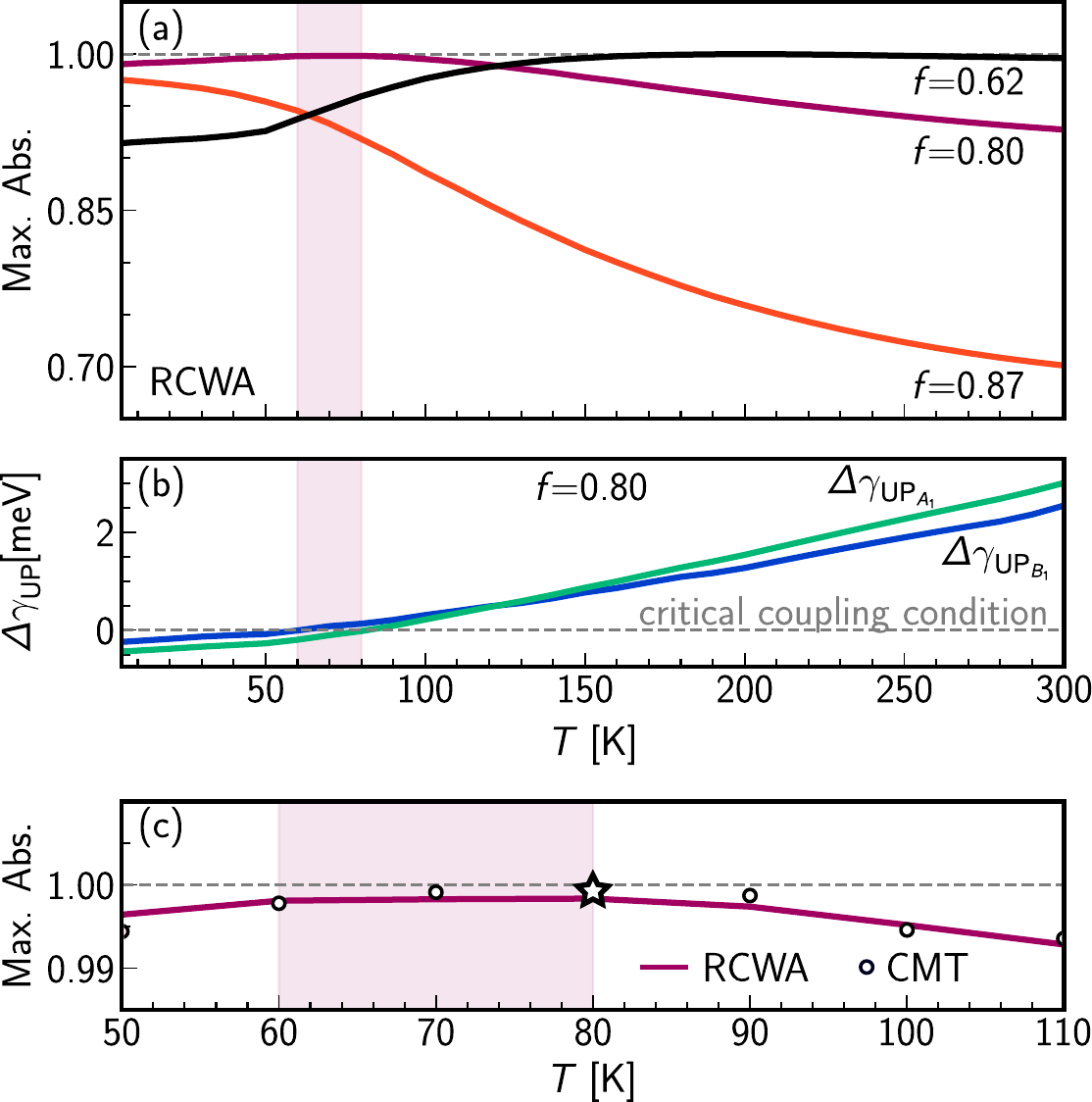}
    \caption{\textbf{(a)} Temperature-dependent maximum absorption of the PhC slab/\ce{WS2} structure for three different PhC filling factors $f = 0.87, \ 0.80,$ and $ 0.62$  calculated with RCWA. Near total absorption for $f=0.80$ and $f=0.62$ can be achieved over a large range of temperatures, as is indicated in the case of $f=0.80$ by the purple shaded region. \textbf{(b)} Critical coupling condition $\Gamma_\text{UP} - \gamma_\text{UP} = 0$ for the two different polariton branches, UP$_{\text{A}_1}$ and UP$_{\text{B}_1}$, resulting from the photonic modes A$_1$ and B$_1$, respectively. The purple shaded region of near total absorption corresponds to the range between the temperatures where either the A or B polariton is critically coupled. \textbf{(c)} Comparison of the maximum absorption between the RCWA (purple line) and the maximum absorption calculated with CMT according to Eq.~(\ref{eq:Elliot}) (black dots) for $f=0.80$. The CMT calculation accurately captures the position of the maximum absorption of $99.8\,\%$ at $80\,$K (indicated by the black star).}
    \label{fig:A_over_T}
\end{figure}

In practical implementations of active semiconductor layers integrated with 2D PhCs, variations in material properties and operating conditions, such as temperature, can significantly modify the excitonic response. However, the geometric tunability of the PhC slab ensures that the degenerate critical coupling condition can be maintained. In the following, we demonstrate how PA can be achieved across a wide range of temperatures by tailoring the PhC filling factor $f$ (the ratio between the silicon area within the unit cell and the total unit cell area.).
The temperature dependence of the system is modeled by a change in the dielectric function of the \ce{WS2} monolayer, as discussed in the Methods section. We assume the dielectric function of \ce{Si} and the geometry of the system to be temperature-independent. 
Figure~\ref{fig:A_over_T}a shows the maximum absorption as a function of temperature for three different PhC filling factors ($f = 0.62, \ 0.80,$ and $0.87$). For each filling factor, the PhC unit cell is scaled such that the crossing of the two photonic modes remains resonant with the exciton. The plotted values correspond to the maximum absorption extracted from full absorption maps that are analogous to Fig.~\ref{fig:RCWA_CMT_Comparison}a. A strong dependence on the filling factor is observed. For $f = 0.87$, PA is not achieved within the considered temperature range. In contrast, for $f = 0.80$, a maximum absorption of $A = 99.8\,\%$ is reached at $80\,$K, while for $f = 0.62$, PA persists even at room temperature ($A = 99.6\,\%$ at $300\,$K). This demonstrates that an appropriate choice of PhC parameters ensures PA remains accessible over a wide temperature range. 
The tunability of PA with temperature and filling factor can be understood by analyzing the decay rates of the polaritons, see Eq.~(\ref{eq:polariton_decay}). 

Figure~\ref{fig:A_over_T}b shows the difference in decay rates $\Delta \gamma_{\text{UP}} = \Gamma_{\text{UP}} - \gamma_{\text{UP}}$ as a function of temperature for $f=0.80$. The green and blue solid lines represent the $\Delta \gamma_{\text{UP}}$ of the UP$_{\text{A}_1}$ and UP$_{\text{B}_1}$, respectively, which cross zero when the critical coupling condition is met. The purple shaded region indicates the temperature window between the points where the individual polaritons are critically coupled, which coincides with the regime of maximum absorption observed in Fig.~\ref{fig:A_over_T}a. Deviations from the critical coupling condition, observed at higher temperatures, lead to reduced absorption, although values exceeding $90\,\%$ are still maintained even at room temperature. The temperature dependence of $\Delta\gamma_\text{UP}$ originates from the increase in the excitonic non-radiative decay rate with increasing temperature. In contrast, the linewidths of the PhC modes remain temperature independent and are modified by the filling factor. Specifically, larger air holes increase scattering of guided modes into the radiative continuum, thus broadening the linewidths of the photonic resonances \cite{suh2003displacement}. 

We determine the optimal temperature of $80\,$K,  at which maximum absorption is achieved, as shown by the black star in Fig.~\ref{fig:A_over_T}c. Here, we find an excellent agreement between the RCWA (purple line) and CMT (black circles) calculations. 
Finally, we assess the robustness of PA under more realistic excitation conditions by modeling the incident light as a finite-width Gaussian beam \cite{kini2020suspended}, which is representative of typical experimental setups. We find that maximum absorption values exceeding $95/97.5\,\%$ are achieved for beam waists larger than $57/75\,\mu$m, corresponding to angular spreads of approximately $2.5/2^\circ$. Further details are provided in the SI Note 3. These results demonstrate that PA is not limited to ideal plane-wave illumination. 
Additionally, our model incorporates the active semiconductor properties exclusively through the dielectric functions. As a result, varying the excitonic decay rates through temperature is effectively equivalent to considering a different excitonic material, since each excitonic resonance in our framework is fully characterized by its spectral position and its radiative and non-radiative decay rates. Consequently, the results presented here are applicable to a broad range of excitonic materials, including other TMDs, 2D perovskites, and magnetic 2D materials.

\section{Discussion}

In this work, we have demonstrated a general mechanism for achieving perfect optical absorption in photonic crystal slabs based on degenerate critical coupling. In the weak coupling regime, we showed that two photonic modes of opposite parity can simultaneously satisfy the critical coupling condition, thereby overcoming the intrinsic $50\,\%$ absorption limit per mode of symmetric two-port systems. By introducing a monolayer excitonic material, we extended the concept of PA via degenerate critical coupling to the strong-coupling regime. At the degeneracy of two polariton branches, near-unity absorption exceeding $99.8\,\%$ was achieved in a photonic crystal slab less than 100\,nm thick. A coupled mode theory was developed to describe the system, accurately reproducing full-wave simulations and providing direct insight into the role of mode parity, coupling strengths, and decay channels. Finally, we demonstrated that this mechanism is adaptable to temperature via tuning of the photonic crystal and tolerant to realistic excitation conditions, highlighting its applicability across a wide range of excitonic materials.

The generality of this approach open several promising directions for future research. Perfect absorption directly implies strong thermal emission \cite{baranov2019nanophotonic}, suggesting applications in tailored thermal emitters and infrared sources. Further optimization of the photonic crystal geometry could enhance robustness with respect to angle and fabrication tolerances. In addition, the strong confinement and efficient energy capture in the polaritonic modes provide an attractive platform for nonlinear optical processes, such as second-harmonic generation. The ability to efficiently populate polariton states is relevant for polariton lasing without population inversion. Finally, the sensitivity of the absorption condition to system parameters could be exploited for sensing applications or photonic logic devices, where small perturbations lead to measurable changes in optical response. 

\section{Methods}
\subsection{Coupled-Mode Theory}

To describe the interaction between photonic modes and excitons, we employ temporal coupled mode theory (CMT) \cite{fan2003temporal}. The dynamics of the photonic and excitonic mode amplitudes, described by the vector $\bm{M}$, and the input/output field amplitudes $\bm{b}^\pm$ are governed by the following equations
\begin{equation}
    \begin{split}
        \hbar \partial_t {\bf M} &= \left( \hbar \underline{\bm{\Omega}} +\underline{\bm{\gamma}} \right) {\bf M} + \underline{\bm{K}}^\top {\bf b^+}\,,\\
        {\bf b^-} &= \underline{\bm{\mathcal{C}}}{\bf b^+} + \underline{\bm{K}} {\bf M} \,,
    \end{split}
    \label{eq:CMT_main}
\end{equation}
where $\underline{\bm{\mathcal{C}}}$ represents the background scattering matrix, accounting for the Fabry-P\'erot modes of the PhC slab and the background permittivity of the TMD monolayer. Furthermore, $\hbar \underline{\bm{\Omega}}$ and $\underline{\bm{\gamma}}$ capture the mode energies and the non-radiative decay rates, respectively. 
The system is modeled in the photon-decoupled regime \cite{balasubrahmaniyam2021coupling}, where each PhC mode couples to a separate excitonic resonance (see SI Note 1). The mode vector is defined as $\bm{M} = (A, X_{\text{A}_1,\text{sc}}, B, X_{\text{B}_1,\text{sc}}, X_\text{wc})^T$, where $A$ and $B$ denote the two photonic modes, and $X_{i,\text{sc}}$ and $X_\text{wc}$ correspond to strongly and weakly coupled excitonic contributions, respectively \cite{kraus2026engineering}. Further details on the CMT are provided in the SI.

\subsection{Temperature Dependence of the Exciton}
While the PhC modes are assumed to be temperature-independent, the excitonic resonances of TMDs—especially the linewidth—are highly temperature-sensitive. To account for this, we model the total excitonic non-radiative decay rate with a phenomenological approach: \cite{meshulam2025temperature,selig2016excitonic}
\begin{equation}
    \gamma_\text{X,tot.}(T) = \gamma_\text{X} + \Gamma(T) = \gamma_\text{X} + \gamma_0 + c_1 T + \frac{c_2}{\exp\left(\frac{\Omega}{k_BT}\right)-1},
\end{equation}
where $\gamma_0$ is the residual scattering rate at zero temperature and $\Omega=20\,$meV represents the average energy of the relevant phonons. The coefficients $c_1$ and $c_2$ describe the linear and superlinear broadening mechanisms, respectively, arising from exciton–phonon interactions. The material-specific parameters are obtained by fitting to experimental temperature-dependent linewidth studies \cite{meshulam2025temperature} (see SI Note 5). 
Furthermore, although the exciton spectral position redshifts with increasing temperature, the PhC geometry can be scaled to tune the photonic mode crossing into resonance with the exciton, meaning this has no qualitative effect on this study.
Lastly, the dielectric function of the \ce{WS2} is modeled by \cite{meshulam2025temperature}:
\begin{equation}
    \varepsilon_{\ce{WS2}}(\omega,T) = \varepsilon_{\text{bg}} - \frac{c}{\omega_\text{X}(T) d_0} \frac{\gamma_\text{X}}{\omega - \omega_\text{X}(T) + i \frac{\Gamma_\text{X}(T)}{2}}\label{eq:eps_WS2}
\end{equation}
where, $c$ is the speed of light in vacuum, $d_0=0.62\,$nm is the monolayer thickness and $\varepsilon_\text{bg} = 16$ is the background permittivity.

\subsection{Rigorous Coupled Wave Analysis}
All simulations shown in this work were performed using rigorous coupled wave analysis (RCWA), a semi-analytical method ideal for efficiently modeling layered structures with one- or two-dimensional in-plane periodicity \cite{whittaker1999scattering}. The method works by expanding the periodic permittivity and electromagnetic fields of each layer into a Fourier series. This approach transforms Maxwell's equations into a matrix eigenvalue problem for each layer, which is solved to find the characteristic modes. Finally, a scattering matrix formalism is used to enforce the boundary conditions of the tangential fields at each interface, leading to the reflection and transmission of the entire device \cite{rumpf2011improved}.
\bibliographystyle{unsrt}
\bibliography{bib}

@article{wang2018colloquium,
  title={Colloquium: Excitons in atomically thin transition metal dichalcogenides},
  author={Wang, Gang and Chernikov, Alexey and Glazov, Mikhail M and Heinz, Tony F and Marie, Xavier and Amand, Thierry and Urbaszek, Bernhard},
  journal={Reviews of Modern Physics},
  volume={90},
  number={2},
  pages={021001},
  year={2018},
  publisher={APS}
}

@article{mueller18,
  title={Exciton physics and device application of two-dimensional transition metal dichalcogenide semiconductors},
  author={Mueller, Thomas and Malic, Ermin},
  journal={njp 2D Materials and Applications},
  volume={2},
  pages={29},
  year={2018},
 }

@article{schneider2018two,
  title={Two-dimensional semiconductors in the regime of strong light-matter coupling},
  author={Schneider, Christian and Glazov, Mikhail M and Korn, Tobias and H{\"o}fling, Sven and Urbaszek, Bernhard},
  journal={Nature communications},
  volume={9},
  number={1},
  pages={2695},
  year={2018},
  publisher={Nature Publishing Group UK London}
}

@article{genco2025femtosecond,
  title={Femtosecond switching of strong light-matter interactions in microcavities with two-dimensional semiconductors},
  author={Genco, Armando and Louca, Charalambos and Cruciano, Cristina and Song, Kok Wee and Trovatello, Chiara and Di Blasio, Giuseppe and Sansone, Giacomo and Randerson, Sam A and Claronino, Peter and Georgiou, Kyriacos and others},
  journal={Nature Communications},
  volume={16},
  number={1},
  pages={6490},
  year={2025},
  publisher={Nature Publishing Group UK London}
}

@article{zhang2018photonic,
  title={Photonic-crystal exciton-polaritons in monolayer semiconductors},
  author={Zhang, Long and Gogna, Rahul and Burg, Will and Tutuc, Emanuel and Deng, Hui},
  journal={Nature communications},
  volume={9},
  number={1},
  pages={713},
  year={2018},
  publisher={Nature Publishing Group UK London}
}

@article{chen2020metasurface,
  title={Metasurface integrated monolayer exciton polariton},
  author={Chen, Yueyang and Miao, Shengnan and Wang, Tianmeng and Zhong, Ding and Saxena, Abhi and Chow, Colin and Whitehead, James and Gerace, Dario and Xu, Xiaodong and Shi, Su-Fei and others},
  journal={Nano Letters},
  volume={20},
  number={7},
  pages={5292--5300},
  year={2020},
  publisher={ACS Publications}
}

@article{fitzgerald2025polariton,
  title={Polariton transport in 2D semiconductors: Phonon-mediated transitions between ballistic, superdiffusive and exciton-limited regimes},
  author={Fitzgerald, Jamie M and Rosati, Roberto and Malic, Ermin},
  journal={Science Advances},
 volume={11},
  pages={46},
  year={2025}
}

@article{fitzgerald2022twist,
  title={Twist angle tuning of moir{\'e} exciton polaritons in van der Waals heterostructures},
  author={Fitzgerald, Jamie M and Thompson, Joshua JP and Malic, Ermin},
  journal={Nano Letters},
  volume={22},
  number={11},
  pages={4468--4474},
  year={2022},
  publisher={ACS Publications}
}

@article{miroshnichenko2010fano,
  title={Fano resonances in nanoscale structures},
  author={Miroshnichenko, Andrey E and Flach, Sergej and Kivshar, Yuri S},
  journal={Reviews of Modern Physics},
  volume={82},
  number={3},
  pages={2257--2298},
  year={2010},
  publisher={APS}
}

@article{fan2003temporal,
  title={Temporal coupled-mode theory for the Fano resonance in optical resonators},
  author={Fan, Shanhui and Suh, Wonjoo and Joannopoulos, John D},
  journal={Journal of the Optical Society of America A},
  volume={20},
  number={3},
  pages={569--572},
  year={2003},
  publisher={Optical Society of America}
}

@article{maggiolini2023strongly,
  title={Strongly enhanced light--matter coupling of monolayer WS2 from a bound state in the continuum},
  author={Maggiolini, Eugenio and Polimeno, Laura and Todisco, Francesco and Di Renzo, Anna and Han, Bo and De Giorgi, Milena and Ardizzone, Vincenzo and Schneider, Christian and Mastria, Rosanna and Cannavale, Alessandro and others},
  journal={Nature Materials},
  volume={22},
  number={8},
  pages={964--969},
  year={2023},
  publisher={Nature Publishing Group UK London}
}

@article{he2023polaritonic,
  title={Polaritonic Chern insulators in monolayer semiconductors},
  author={He, Li and Wu, Jingda and Jin, Jicheng and Mele, Eugene J and Zhen, Bo},
  journal={Physical Review Letters},
  volume={130},
  number={4},
  pages={043801},
  year={2023},
  publisher={APS}
}

@article{kravtsov2020nonlinear,
  title={Nonlinear polaritons in a monolayer semiconductor coupled to optical bound states in the continuum},
  author={Kravtsov, Vasily and Khestanova, Ekaterina and Benimetskiy, Fedor A and Ivanova, Tatiana and Samusev, Anton K and Sinev, Ivan S and Pidgayko, Dmitry and Mozharov, Alexey M and Mukhin, Ivan S and Lozhkin, Maksim S and others},
  journal={Light: Science \& Applications},
  volume={9},
  number={1},
  pages={56},
  year={2020},
  publisher={Nature Publishing Group UK London}
}

@article{whittaker1999scattering,
  title={Scattering-matrix treatment of patterned multilayer photonic structures},
  author={Whittaker, DM and Culshaw, IS},
  journal={Physical Review B},
  volume={60},
  number={4},
  pages={2610},
  year={1999},
  publisher={APS}
}

@article{baranov2017coherent,
  title={Coherent perfect absorbers: linear control of light with light},
  author={Baranov, Denis G and Krasnok, Alex and Shegai, Timur and Al{\`u}, Andrea and Chong, Yidong},
  journal={Nature Reviews Materials},
  volume={2},
  number={12},
  pages={1--14},
  year={2017},
  publisher={Nature Publishing Group}
}

@article{chong2010coherent,
  title={Coherent perfect absorbers: time-reversed lasers},
  author={Chong, YD and Ge, Li and Cao, Hui and Stone, A Douglas},
  journal={Physical review letters},
  volume={105},
  number={5},
  pages={053901},
  year={2010},
  publisher={APS}
}

@article{wan2011time,
  title={Time-reversed lasing and interferometric control of absorption},
  author={Wan, Wenjie and Chong, Yidong and Ge, Li and Noh, Heeso and Stone, A Douglas and Cao, Hui},
  journal={Science},
  volume={331},
  number={6019},
  pages={889--892},
  year={2011},
  publisher={American Association for the Advancement of Science}
}

@article{noh2012perfect,
  title={Perfect coupling of light to surface plasmons by coherent absorption},
  author={Noh, Heeso and Chong, Yidong and Stone, A Douglas and Cao, Hui},
  journal={Physical review letters},
  volume={108},
  number={18},
  pages={186805},
  year={2012},
  publisher={APS}
}

@article{jimenez2016ultra,
  title={Ultra-thin metamaterial for perfect and quasi-omnidirectional sound absorption},
  author={Jim{\'e}nez, Noe and Huang, Weichun and Romero-Garc{\'\i}a, Vicent and Pagneux, Vincent and Groby, J-P},
  journal={Applied Physics Letters},
  volume={109},
  number={12},
  year={2016},
  publisher={AIP Publishing}
}

@article{kishino1991resonant,
  title={Resonant cavity-enhanced (RCE) photodetectors},
  author={Kishino, Katsumi and Unlu, M Selim and Chyi, J-I and Reed, J and Arsenault, L and Morkoc, Hadis},
  journal={IEEE Journal of Quantum Electronics},
  volume={27},
  number={8},
  pages={2025--2034},
  year={1991},
  publisher={IEEE}
}

@article{jin2021organic,
  title={Organic sub-bandgap Schottky barrier photodetectors with near-infrared coherent perfect absorption},
  author={Jin, Yeonghoon and Kim, Hyung Suk and Park, Junghoon and Yoo, Seunghyup and Yu, Kyoungsik},
  journal={ACS Photonics},
  volume={8},
  number={9},
  pages={2618--2625},
  year={2021},
  publisher={ACS Publications}
}

@article{polman2012photonic,
  title={Photonic design principles for ultrahigh-efficiency photovoltaics},
  author={Polman, Albert and Atwater, Harry A},
  journal={Nature materials},
  volume={11},
  number={3},
  pages={174--177},
  year={2012},
  publisher={Nature Publishing Group UK London}
}

@article{liew2016coherent,
  title={Coherent control of photocurrent in a strongly scattering photoelectrochemical system},
  author={Liew, Seng Fatt and Popoff, S\'{e}bastien M and Sheehan, Stafford W and Goetschy, Arthur and Schmuttenmaer, Charles A and Stone, A Douglas and Cao, Hui},
  journal={ACS photonics},
  volume={3},
  number={3},
  pages={449--455},
  year={2016},
  publisher={ACS Publications}
}

@article{watts2012metamaterial,
  title={Metamaterial electromagnetic wave absorbers},
  author={Watts, Claire M and Liu, Xianliang and Padilla, Willie J},
  journal={Advanced materials},
  volume={24},
  number={23},
  pages={OP98--OP120},
  year={2012},
  publisher={Wiley Online Library}
}

@article{perea2022exciton,
  title={Exciton optics, dynamics, and transport in atomically thin semiconductors},
  author={Perea-Causin, Raul and Erkensten, Daniel and Fitzgerald, Jamie M and Thompson, Joshua JP and Rosati, Roberto and Brem, Samuel and Malic, Ermin},
  journal={APL Materials},
  volume={10},
  number={10},
  year={2022},
  publisher={AIP Publishing}
}

@article{slobodkin2022massively,
  title={Massively degenerate coherent perfect absorber for arbitrary wavefronts},
  author={Slobodkin, Yevgeny and Weinberg, Gil and H{\"o}rner, Helmut and Pichler, Kevin and Rotter, Stefan and Katz, Ori},
  journal={Science},
  volume={377},
  number={6609},
  pages={995--998},
  year={2022},
  publisher={American Association for the Advancement of Science}
}

@article{piper2014total,
  title={Total absorption by degenerate critical coupling},
  author={Piper, Jessica R and Liu, Victor and Fan, Shanhui},
  journal={Applied Physics Letters},
  volume={104},
  number={25},
  year={2014},
  publisher={AIP Publishing}
}

@article{horng2020perfect,
  title={Perfect absorption by an atomically thin crystal},
  author={Horng, Jason and Martin, Eric W and Chou, Yu-Hsun and Courtade, Emmanuel and Chang, Tsu-chi and Hsu, Chu-Yuan and Wentzel, Michael-Henr and Ruth, Hanna G and Lu, Tien-chang and Cundiff, Steven T and others},
  journal={Physical Review Applied},
  volume={14},
  number={2},
  pages={024009},
  year={2020},
  publisher={APS}
}

@article{lee2023achieving,
  title={Achieving near-perfect light absorption in atomically thin transition metal dichalcogenides through band nesting},
  author={Lee, Seungjun and Seo, Dongjea and Park, Sang Hyun and Izquierdo, Nezhueytl and Lee, Eng Hock and Younas, Rehan and Zhou, Guanyu and Palei, Milan and Hoffman, Anthony J and Jang, Min Seok and others},
  journal={Nature communications},
  volume={14},
  number={1},
  pages={3889},
  year={2023},
  publisher={Nature Publishing Group UK London}
}

@article{epstein2020near,
  title={Near-unity light absorption in a monolayer WS2 van der Waals heterostructure cavity},
  author={Epstein, Itai and Terres, Bernat and Chaves, Andre J and Pusapati, Varun-Varma and Rhodes, Daniel A and Frank, Bettina and Zimmermann, Valentin and Qin, Ying and Watanabe, Kenji and Taniguchi, Takashi and others},
  journal={Nano letters},
  volume={20},
  number={5},
  pages={3545--3552},
  year={2020},
  publisher={ACS Publications}
}

@article{horng2019engineering,
  title={Engineering radiative coupling of excitons in 2D semiconductors},
  author={Horng, Jason and Chou, Yu-Hsun and Chang, Tsu-Chi and Hsu, Chu-Yuan and Lu, Tien-Chang and Deng, Hui},
  journal={Optica},
  volume={6},
  number={11},
  pages={1443--1448},
  year={2019},
  publisher={Optical Society of America}
}

@article{thongrattanasiri2012complete,
  title={Complete optical absorption in periodically patterned graphene},
  author={Thongrattanasiri, Sukosin and Koppens, Frank HL and Garc{\'\i}a de Abajo, F Javier},
  journal={Physical review letters},
  volume={108},
  number={4},
  pages={047401},
  year={2012},
  publisher={APS}
}

@article{jariwala2016near,
  title={Near-unity absorption in van der Waals semiconductors for ultrathin optoelectronics},
  author={Jariwala, Deep and Davoyan, Artur R and Tagliabue, Giulia and Sherrott, Michelle C and Wong, Joeson and Atwater, Harry A},
  journal={Nano letters},
  volume={16},
  number={9},
  pages={5482--5487},
  year={2016},
  publisher={ACS Publications}
}

@article{canales2023perfect,
  title={Perfect absorption and strong coupling in supported MoS2 multilayers},
  author={Canales, Adriana and Kotov, Oleg and Shegai, Timur O},
  journal={ACS nano},
  volume={17},
  number={4},
  pages={3401--3411},
  year={2023},
  publisher={ACS Publications}
}

@article{li2019engineering,
  title={Engineering optical absorption in graphene and other 2D materials: advances and applications},
  author={Li, Qiang and Lu, Jun and Gupta, Prince and Qiu, Min},
  journal={Advanced Optical Materials},
  volume={7},
  number={20},
  pages={1900595},
  year={2019},
  publisher={Wiley Online Library}
}

@article{galiffi2026optical,
  title={Optical coherent perfect absorption and amplification in a time-varying medium},
  author={Galiffi, Emanuele and Harwood, Anthony C and Vezzoli, Stefano and Tirole, Romain and Al{\`u}, Andrea and Sapienza, Riccardo},
  journal={Nature Photonics},
  pages={1--7},
  year={2026},
  publisher={Nature Publishing Group UK London}
}

@article{piper2014total_2,
  title={Total absorption in a graphene monolayer in the optical regime by critical coupling with a photonic crystal guided resonance},
  author={Piper, Jessica R and Fan, Shanhui},
  journal={Acs Photonics},
  volume={1},
  number={4},
  pages={347--353},
  year={2014},
  publisher={ACS Publications}
}

@article{rao2014coherent,
  title={Coherent control of light interaction with graphene},
  author={Rao, Shraddha M and Heitz, Julius JF and Roger, Thomas and Westerberg, Niclas and Faccio, Daniele},
  journal={Optics letters},
  volume={39},
  number={18},
  pages={5345--5347},
  year={2014},
  publisher={Optical Society of America}
}

@article{zhang2012controlling,
  title={Controlling light-with-light without nonlinearity},
  author={Zhang, Jianfa and MacDonald, Kevin F and Zheludev, Nikolay I},
  journal={Light: Science \& Applications},
  volume={1},
  number={7},
  pages={e18--e18},
  year={2012},
  publisher={Nature Publishing Group}
}

@article{zanotto2014perfect,
  title={Perfect energy-feeding into strongly coupled systems and interferometric control of polariton absorption},
  author={Zanotto, Simone and Mezzapesa, Francesco P and Bianco, Federica and Biasiol, Giorgio and Baldacci, Lorenzo and Vitiello, Miriam Serena and Sorba, Lucia and Colombelli, Raffaele and Tredicucci, Alessandro},
  journal={Nature Physics},
  volume={10},
  number={11},
  pages={830--834},
  year={2014},
  publisher={Nature Publishing Group UK London}
}

@article{zhou2016perfect,
  title={Perfect single-sided radiation and absorption without mirrors},
  author={Zhou, Hengyun and Zhen, Bo and Hsu, Chia Wei and Miller, Owen D and Johnson, Steven G and Joannopoulos, John D and Solja{\v{c}}i{\'c}, Marin},
  journal={Optica},
  volume={3},
  number={10},
  pages={1079--1086},
  year={2016},
  publisher={Optical Society of America}
}

@article{baranov2019nanophotonic,
  title={Nanophotonic engineering of far-field thermal emitters},
  author={Baranov, Denis G and Xiao, Yuzhe and Nechepurenko, Igor A and Krasnok, Alex and Al{\`u}, Andrea and Kats, Mikhail A},
  journal={Nature materials},
  volume={18},
  number={9},
  pages={920--930},
  year={2019},
  publisher={Nature Publishing Group UK London}
}

@article{kats2016optical,
  title={Optical absorbers based on strong interference in ultra-thin films},
  author={Kats, Mikhail A and Capasso, Federico},
  journal={Laser \& Photonics Reviews},
  volume={10},
  number={5},
  pages={735--749},
  year={2016},
  publisher={Wiley Online Library}
}

@article{euve2023perfect,
  title={Perfect resonant absorption of guided water waves by Autler-Townes splitting},
  author={Euv{\'e}, L-P and Pham, Kim and Porter, Richard and Petitjeans, Philippe and Pagneux, Vincent and Maurel, Agn{\`e}s},
  journal={Physical Review Letters},
  volume={131},
  number={20},
  pages={204002},
  year={2023},
  publisher={APS}
}

@article{balasubrahmaniyam2021coupling,
  title={Coupling and decoupling of polaritonic states in multimode cavities},
  author={Balasubrahmaniyam, M and Genet, Cyriaque and Schwartz, Tal},
  journal={Physical Review B},
  volume={103},
  number={24},
  pages={L241407},
  year={2021},
  publisher={APS}
}

@article{georgiou2021observation,
  title={Observation of photon-mode decoupling in a strongly coupled multimode microcavity},
  author={Georgiou, Kyriacos and McGhee, Kirsty E and Jayaprakash, Rahul and Lidzey, David G},
  journal={The Journal of Chemical Physics},
  volume={154},
  number={12},
  year={2021},
  publisher={AIP Publishing}
}

@article{li2014equivalent,
  title={An equivalent realization of coherent perfect absorption under single beam illumination},
  author={Li, Sucheng and Luo, Jie and Anwar, Shahzad and Li, Shuo and Lu, Weixin and Hang, Zhi Hong and Lai, Yun and Hou, Bo and Shen, Mingrong and Wang, Chinhua},
  journal={Scientific reports},
  volume={4},
  number={1},
  pages={7369},
  year={2014},
  publisher={Nature Publishing Group UK London}
}

@article{piper2016broadband,
  title={Broadband absorption enhancement in solar cells with an atomically thin active layer},
  author={Piper, Jessica R and Fan, Shanhui},
  journal={Acs Photonics},
  volume={3},
  number={4},
  pages={571--577},
  year={2016},
  publisher={ACS Publications}
}

@article{thareja2015electrically,
  title={Electrically tunable coherent optical absorption in graphene with ion gel},
  author={Thareja, Vrinda and Kang, Ju-Hyung and Yuan, Hongtao and Milaninia, Kaveh M and Hwang, Harold Y and Cui, Yi and Kik, Pieter G and Brongersma, Mark L},
  journal={Nano letters},
  volume={15},
  number={3},
  pages={1570--1576},
  year={2015},
  publisher={ACS Publications}
}

@article{zhang2014coherent,
  title={Coherent perfect absorption and transparency in a nanostructured graphene film},
  author={Zhang, Jianfa and Guo, Chucai and Liu, Ken and Zhu, Zhihong and Ye, Weimin and Yuan, Xiaodong and Qin, Shiqiao},
  journal={Optics express},
  volume={22},
  number={10},
  pages={12524--12532},
  year={2014},
  publisher={Optical Society of America}
}

@article{konig2025magneto,
  title={Magneto-optics of anisotropic exciton polaritons in two-dimensional perovskites},
  author={K\"onig, Jonas K and Fitzgerald, Jamie M and Malic, Ermin},
  journal={Nano Letters},
  volume={25},
  number={21},
  pages={8519--8526},
  year={2025},
  publisher={ACS Publications}
}

@article{mondal2025switching,
  title={Switching polariton screening in MoS2 microcavity toward polaritonics},
  author={Mondal, Ashok and Biswas, Chandan and Ghising, Pramod and Moon, Byoung Hee and Kim, Ki Kang and Lee, Young Hee},
  journal={Science Advances},
  volume={11},
  number={8},
  pages={eadr7202},
  year={2025},
  publisher={American Association for the Advancement of Science}
}

@article{gu2023polaritonic,
  title={Polaritonic coherent perfect absorption based on self-hybridization of a quasi-bound state in the continuum and exciton},
  author={Gu, Xin and Liu, Xing and Yan, Xiao-Fei and Du, Wen-Juan and Lin, Qi and Wang, Ling-Ling and Liu, Gui-Dong},
  journal={Optics Express},
  volume={31},
  number={3},
  pages={4691--4700},
  year={2023},
  publisher={Optica Publishing Group}
}

@article{liu2017experimental,
  title={Experimental realization of a terahertz all-dielectric metasurface absorber},
  author={Liu, Xinyu and Fan, Kebin and Shadrivov, Ilya V and Padilla, Willie J},
  journal={Optics express},
  volume={25},
  number={1},
  pages={191--201},
  year={2017},
  publisher={Optical Society of America}
}

@article{kraus2026engineering,
      title={Engineering strong coupling in ultra-compact photonic crystal/2D material platforms}, 
      author={Eleonora P. Kraus and Jamie M. Fitzgerald and Carlos Maciel-Escudero and Ermin Malic},
      year={2026},
      eprint={2604.12779},
      journal={arXiv},
      archivePrefix={arXiv},
      primaryClass={physics.optics},
      url={https://arxiv.org/abs/2604.12779}, 
}

@article{kini2020suspended,
  title={Suspended photonic crystal membranes in AlGaAs heterostructures for integrated multi-element optomechanics},
  author={Kini Manjeshwar, Sushanth and Elkhouly, Karim and Fitzgerald, Jamie M and Ekman, Martin and Zhang, Yanchao and Zhang, Fan and Wang, Shu Min and Tassin, Philippe and Wieczorek, Witlef},
  journal={Applied Physics Letters},
  volume={116},
  number={26},
  year={2020},
  publisher={AIP Publishing}
}

@article{selig2016excitonic,
  title={Excitonic linewidth and coherence lifetime in monolayer transition metal dichalcogenides},
  author={Selig, Malte and Bergh{\"a}user, Gunnar and Raja, Archana and Nagler, Philipp and Sch{\"u}ller, Christian and Heinz, Tony F and Korn, Tobias and Chernikov, Alexey and Malic, Ermin and Knorr, Andreas},
  journal={Nature communications},
  volume={7},
  number={1},
  pages={13279},
  year={2016},
  publisher={Nature Publishing Group UK London}
}

@article{aspnes1983dielectric,
  title = {Dielectric functions and optical parameters of Si, Ge, GaP, GaAs, GaSb, InP, InAs, and InSb from 1.5 to 6.0 eV},
  author = {Aspnes, D. E. and Studna, A. A.},
  journal = {Phys. Rev. B},
  volume = {27},
  issue = {2},
  pages = {985--1009},
  numpages = {0},
  year = {1983},
  month = {Jan},
  publisher = {American Physical Society},
  doi = {10.1103/PhysRevB.27.985},
  url = {https://link.aps.org/doi/10.1103/PhysRevB.27.985}
}

@article{meshulam2025temperature,
author = {Meshulam, Matan and Kahlon, Anabel Atash and Gershuni, Yonatan and Poirier, Thomas and Edgar, James H. and Tongay, Seth Ariel and Epstein, Itai},
title = {Temperature-Dependent Optical and Polaritonic Properties of Excitons in hBN-Encapsulated Monolayer TMDs},
journal = {Advanced Optical Materials},
volume = {14},
number = {7},
pages = {e02535},
doi = {https://doi.org/10.1002/adom.202502535},
url = {https://advanced.onlinelibrary.wiley.com/doi/abs/10.1002/adom.202502535},
eprint = {https://advanced.onlinelibrary.wiley.com/doi/pdf/10.1002/adom.202502535},
year = {2026}
}

@article{rumpf2011improved,
  title={Improved formulation of scattering matrices for semi-analytical methods that is consistent with convention},
  author={Rumpf, Raymond C},
  journal={Progress In Electromagnetics Research B},
  volume={35},
  pages={241--261},
  year={2011},
  publisher={EMW Publishing}
}

@article{baffou2013thermo,
  title={Thermo-plasmonics: using metallic nanostructures as nano-sources of heat},
  author={Baffou, Guillaume and Quidant, Romain},
  journal={Laser \& Photonics Reviews},
  volume={7},
  number={2},
  pages={171--187},
  year={2013},
  publisher={Wiley Online Library}
}

@article{lalanne1997depth,
author = {Philippe Lalanne and Dominique Lemercier-Lalanne},
journal = {J. Opt. Soc. Am. A},
number = {2},
pages = {450--459},
publisher = {Optica Publishing Group},
title = {Depth dependence of the effective properties of subwavelength gratings},
volume = {14},
year = {1997},
url = {https://opg.optica.org/josaa/abstract.cfm?URI=josaa-14-2-450},
doi = {10.1364/JOSAA.14.000450}
}

@article{chen2017high,
  title={High-finesse Fabry--Perot cavities with bidimensional Si3N4 photonic-crystal slabs},
  author={Chen, Xu and Chardin, Cl{\'e}ment and Makles, Kevin and Ca{\"e}r, Charles and Chua, Sheon and Braive, R{\'e}my and Robert-Philip, Isabelle and Briant, Tristan and Cohadon, Pierre-Fran{\c{c}}ois and Heidmann, Antoine and others},
  journal={Light: Science \& Applications},
  volume={6},
  number={1},
  pages={e16190--e16190},
  year={2017},
  publisher={Nature Publishing Group}
}

@article{ciers2017propagating,
  title={Propagating polaritons in III-nitride slab waveguides},
  author={Ciers, Joachim and Roch, Jonas G and Carlin, J-F and Jacopin, Gw{\'e}nol{\'e} and Butt{\'e}, Rapha{\"e}l and Grandjean, Nicolas},
  journal={Physical Review Applied},
  volume={7},
  number={3},
  pages={034019},
  year={2017},
  publisher={APS}
}

@article{tikhodeev2002quasiguided,
  title={Quasiguided modes and optical properties of photonic crystal slabs},
  author={Tikhodeev, Sergei G and Yablonskii, AL and Muljarov, EA and Gippius, Nikolay A and Ishihara, Teruya},
  journal={Physical Review B},
  volume={66},
  number={4},
  pages={045102},
  year={2002},
  publisher={APS}
}

@article{quaranta2018recent,
  title={Recent advances in resonant waveguide gratings},
  author={Quaranta, Giorgio and Basset, Guillaume and Martin, Olivier JF and Gallinet, Benjamin},
  journal={Laser \& Photonics Reviews},
  volume={12},
  number={9},
  pages={1800017},
  year={2018},
  publisher={Wiley Online Library}
}

@book{yehoptical,
  title={Optical Waves in Layered Media},
  author={Yeh, P},
  year={1988},
  publisher={ J. Wiley and Sons}
}

@article{suh2004temporal,
  title={Temporal coupled-mode theory and the presence of non-orthogonal modes in lossless multimode cavities},
  author={Suh, Wonjoo and Wang, Zheng and Fan, Shanhui},
  journal={IEEE Journal of Quantum Electronics},
  volume={40},
  number={10},
  pages={1511--1518},
  year={2004},
  publisher={IEEE}
}

@article{suh2003displacement,
  title={Displacement-sensitive photonic crystal structures based on guided resonance in photonic crystal slabs},
  author={Suh, Wonjoo and Yanik, MF and Solgaard, Olav and Fan, Shanhui},
  journal={Applied physics letters},
  volume={82},
  number={13},
  pages={1999--2001},
  year={2003},
  publisher={American Institute of Physics}
}
\end{document}

\title{Supplementary Information: Perfect Absorption in the Strong Coupling Regime via Degenerate Critical Coupling}
\author{Eleonora P. Kraus} 
\email{eleonora.kraus@physik.uni-marburg.de} 
\affiliation{Department of Physics, Philipps-Universit\"{a}t Marburg, 35037 Marburg, Germany} 
\affiliation{mar.quest|Marburg Center for Quantum Materials and Sustainable Technologies, 35032 Marburg, Germany} 
\author{Jamie M. Fitzgerald} 
\affiliation{Department of Physics, Philipps-Universit\"{a}t Marburg, 35037 Marburg, Germany} 
\affiliation{mar.quest|Marburg Center for Quantum Materials and Sustainable Technologies, 35032 Marburg, Germany} 
\author{Carlos Maciel-Escudero} 
\affiliation{Department of Physics, Philipps-Universit\"{a}t Marburg, 35037 Marburg, Germany} 
\affiliation{mar.quest|Marburg Center for Quantum Materials and Sustainable Technologies, 35032 Marburg, Germany} 
\author{Ermin Malic} 
\affiliation{Department of Physics, Philipps-Universit\"{a}t Marburg, 35037 Marburg, Germany} 
\affiliation{mar.quest|Marburg Center for Quantum Materials and Sustainable Technologies, 35032 Marburg, Germany}

\maketitle
\sisection{Photon Decoupled Model and Zone Folding of the Polaritons}

\begin{figure}[t!]
    \centering
    \includegraphics[width=0.85\linewidth]{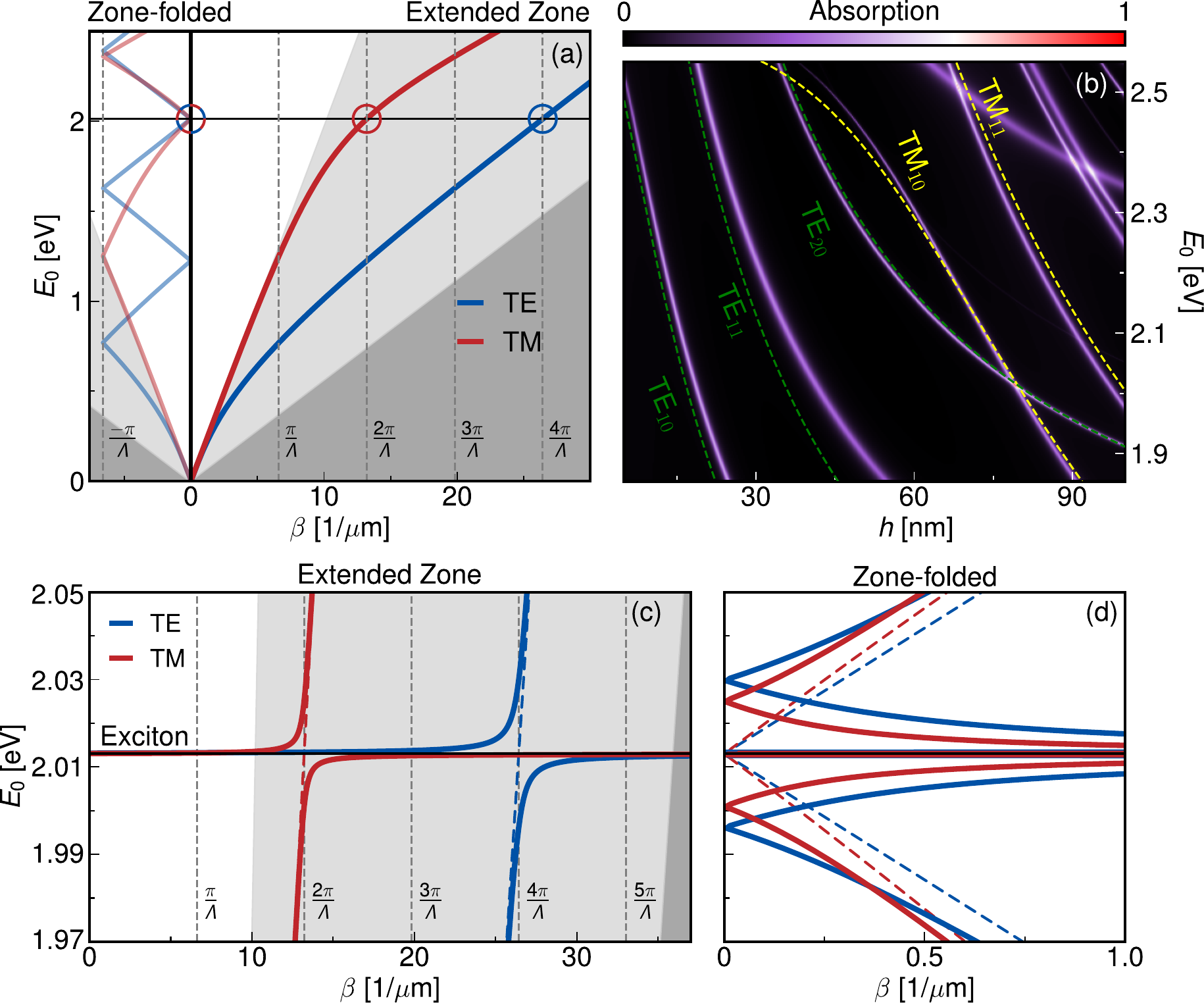}
    \caption{\textbf{(a)} Dispersion relations for the fundamental TE (blue) and TM (red) guided modes of a slab waveguide of thickness $79.9\,$nm and an effective dielectric constant set by Eq.~\ref{eq:effective_eps} with the same PhC lattice parameters as Fig.~2 in the main text. The right side shows the extended zone scheme (dark waveguide modes), while the left side shows the corresponding zone-folded bands (bright guided mode resonances). The light grey shaded region denotes the silicon lightcone that supports slab waveguide modes: $E_0/(c\hbar) < \beta < n_{\text{Si}}E_0/(c \hbar)$. Vertical dashed lines mark the Brillouin zone boundaries at multiples of $\pi/\Lambda$. \textbf{(b)} A zoom in of the RCWA-calculated absorption map versus energy and slab thickness from Fig.~2a in the main text. Overlaid are the calculated guided mode resonances from the effective slab model. \textbf{(c)} Extended zone dispersion for the PhC--TMD system in the vicinity of the exciton resonance. The solid lines show strong coupling (anti-crossing) between the waveguide modes (dashed lines) and the flat exciton band (black horizontal line), forming a pair of upper and lower polariton branches for TE and TM. \textbf{(d)} The corresponding zone-folded dispersion relations for the strongly coupled PhC--TMD system depicted in panel c, illustrating four polariton branches and an uncoupled exciton branch within the lightcone.}
    \label{fig:SI_Photon_Coupled}
\end{figure}

In order to characterize the underlying photonic modes and build intuition for light-matter coupling in photonic crystal (PhC) systems, it is instructive to use an effective slab waveguide model combined with an empty-lattice approximation \cite{tikhodeev2002quasiguided}, which we hereafter refer to as the \emph{effective slab model}. This approach, detailed in our previous work \cite{kraus2026engineering} and elsewhere \cite{chen2017high,quaranta2018recent}, models the bare PhC slab as an effective homogeneous dielectric slab supporting both transverse electric (TE) and magnetic (TM) waveguide modes, see the blue and red curves in Fig.~\ref{fig:SI_Photon_Coupled}a within the ``extended zone'' scheme. The effective slab waveguide is described using a volume-averaged dielectric constant \cite{lalanne1997depth}: 
\begin{equation}
\varepsilon_{\text{eff}} = \varepsilon_{\text{Si}}f + \varepsilon_{\text{Air}}(1-f). \label{eq:effective_eps}
\end{equation}
The grating periodicity in each in-plane direction, $\Lambda_i \ (i=x,y)$, is treated as a periodic perturbation that imparts an in-plane momentum of $m_i2\pi/\Lambda_i$. This momentum transfer couples slab waveguide modes to the far-field, i.e., dark guided modes are zone-folded into the lightcone to form bright (leaky) guided-mode resonances \cite{quaranta2018recent}. This process is illustrated by the fainter blue and red curves on the left side of Fig.~\ref{fig:SI_Photon_Coupled}a within the ``zone-folded'' scheme. In this work we predominantly focus on photonic modes excited at normal incidence ($\beta=0$). These can be found by solving the following equation for the energy $E$ \cite{chen2017high}:
\begin{align}
    0 &= 1 + r_{12}r_{23} e^{2i k_z^{(2)} h}, \label{eq:phase_condition} \\ 
    k_z^{(2)} &=\sqrt{  \left(\frac{E}{ \hbar c}\right)^2 \varepsilon_{\text{eff}} - \left( m_x \frac{2\pi}{\Lambda_x} \right)^2 - \left( m_y \frac{2\pi}{\Lambda_y} \right)^2}, 
\end{align}
which is a condition on the round-trip phase accumulation in the out-of-plane direction for a propagating guided mode in the slab. Here, $h$ is the effective slab waveguide thickness, $k_z^{(2)}$ is the out-of-plane momentum component within the effective slab, and $r_{ij}$ are the Fresnel coefficients at the interface between media $i$ and $j$. For TE and TM polarizations, these are defined as:
\begin{align}
    r_{ij}^{\text{TE}}&= \frac{k_z^{(i)} - k_z^{(j)}}{k_z^{(i)} + k_z^{(j)}}, \\
    r_{ij}^{\text{TM}}&=\frac{\varepsilon_i k_z^{(j)} - \varepsilon_j k_z^{(i)}}{\varepsilon_i k_z^{(j)} + \varepsilon_j k_z^{(i)}}. \label{eq:TM_Fresnel}
\end{align}

The resulting mode energies are purely real as there is no radiative decay or loss included within the model. By solving Eqs.~\ref{eq:phase_condition}--\ref{eq:TM_Fresnel} for different slab thicknesses, we can characterize the PhC modes appearing as distinct absorption bands in the spectra shown in Fig.~2a of the main text. For moderate energies and small slab thicknesses, each mode can be identified by a specific number of zone foldings along the $x$ and $y$ directions, corresponding to integer multiples of $2\pi/\Lambda_i$ imparted by the PhC periodicity. We restrict our attention to the fundamental TE and TM waveguide modes (for the fundamental TE/TM modes, the transverse electric/magnetic field profile has no nodes within the waveguide), which is valid for thin slabs. A magnified view of the relevant energy and thickness range is shown in Fig.~\ref{fig:SI_Photon_Coupled}b. We find excellent qualitative agreement between the absorption calculated via rigorous coupled-wave analysis (RCWA) and the predicted mode positions. In particular, this allows us to identify the even A$_1$ mode introduced in the main text as the TE$_{20}$ mode (i.e., a TE waveguide mode at $\beta=4\pi/\Lambda$ that has been zone-folded twice times along either the $x$ or $y$ direction), and the odd B$_1$ mode as the TM$_{10}$ mode (i.e., a TM waveguide mode at $\beta=2\pi/\Lambda$ that has been zone-folded once along either the $x$ or $y$ direction). This matches the well-known electric field configurations for the fundamental waveguide modes of a symmetric slab. Specifically, across the slab mid-plane, the transverse (in-plane) electric field possesses even parity for the fundamental TE mode and odd parity for the fundamental TM mode \cite{yehoptical}.

The effective slab model also provides insight into engineering the PhC to achieve degenerate critical coupling: the underlying TE and TM waveguide modes must cross the target energy at momenta separated exactly by an integer multiple of $2\pi/\Lambda$. This is represented graphically for a specific slab thickness of $h=79.9$ nm in Fig.~\ref{fig:SI_Photon_Coupled}a. Here,  the red (TM) and blue (TE) circles show the intersection of the respective waveguide modes with the exciton energy in the extended zone scheme. For this particular geometry, the zone-folding of these modes results in two degenerate guided-mode resonances of opposite parity at $\beta=0$. Note that, in general, the empty-lattice approximation will not capture bandgap openings at the Brillouin zone center and edges. The grating periodicity is treated purely as a mechanism for momentum transfer, neglecting the coupling between different PhC modes. Interestingly, in this particular case, the opposite parity of the photonic modes forbids any interaction, resulting in excellent agreement with the simple effective slab model in Fig.~\ref{fig:SI_Photon_Coupled}b.

The effective slab model can also be applied to the strong coupling regime, as shown in Fig.~\ref{fig:SI_Photon_Coupled}c. Here, the two waveguide modes (dashed curves) couple with the 1s exciton resonance of the \ce{WS2} monolayer (horizontal black line) to form dark waveguide polaritons. This interaction leads to a pair of lower and upper polariton branches for both TE and TM modes (blue and red solid curves, respectively). The coupling strength can be estimated using \cite{ciers2017propagating, kraus2026engineering}
\begin{equation}
    \label{eq:g}
    g= \hbar \mathcal{F} \sqrt{\frac{e^2f}{m_\text{e}\varepsilon_0\varepsilon_{\text{eff}}L_{\text{mode}}}},
\end{equation}
where $f$ is the exciton oscillator strength per unit area and $L_{\text{mode}}$ is the effective slab waveguide mode length. The factor $\mathcal{F}$ accounts for the displacement of the TMD monolayer away from the electric field maximum. It is defined as the ratio between the electric field maximum and that at the position of the monolayer.  From Eq.~\ref{eq:g}, we obtain Rabi splittings of $15.71$ and $11.22\,$meV, for the TE and TM modes respectively. These values are in excellent agreement with the Rabi splittings $16.86$ and $12.05\,$meV extracted from a coupled mode theory (CMT) fit to full RCWA calculations.

% Note that it is important to normalize the total electric field, in particular taking into account the $z$ component of the TM field.

As with the bare PhC case, a zone-folded scheme can be used to explore how the dark waveguide polaritons are brightened by the PhC periodicity. The resulting polariton branches are shown in Fig.~\ref{fig:SI_Photon_Coupled}d. Crucially, the model predicts the correct number of polariton branches: two lower and two upper branches of opposite parity (i.e., there is a TE and TM lower polariton branch, likewise for the upper polariton). There is also a collection of flat branches at the exciton energy, which arise from low-momentum bright excitons (corresponding to uncoupled excitons in the effective slab model) as well as the guided polariton dispersions folded from neighboring Brillouin zones (see the SI of Ref.~\cite{kraus2026engineering}). The effective slab model also provides insight into correctly constructing a CMT for exciton polaritons in PhC systems. Figures~\ref{fig:SI_Photon_Coupled}a and c clearly highlight that distinct exciton momentum states couple to the TE and TM waveguide modes, justifying the inclusion of two different exciton states in Eqs.~\ref{eq:CMT_hbarOmega} and \ref{eq:CMT_gamma} of the CMT (see Supplementary Note \ref{sisec:CMT}). Furthermore, as each of these distinct exciton states couples to only one of the two waveguide modes, this explains the necessity of a photon-decoupled Hamiltonian description for strong coupling in PhC systems. Additionally, the weak coupling exciton state included in Eqs.~\ref{eq:CMT_hbarOmega} and \ref{eq:CMT_gamma} corresponds to a bright exciton state within the lightcone that interacts directly with free-space photons.

\sisection{Coupled-Mode Theory for Photonic Crystal-TMD Strong Coupling}\label{sisec:CMT}
To describe the interaction between photonic modes and excitons, we employ a temporal coupled mode theory (CMT). The system is modeled in the photon-decoupled approach, as discussed in Supplementary Note 1. The dynamics of the amplitudes of the two photonic modes $A_1$ and $B_1$, interacting with both the direct and zone-folded finite-momentum excitons, is determined by the following equations \cite{fan2003temporal}:
\begin{equation}
    \begin{split}
        \hbar \partial_t {\bf M} &= \left( \hbar \underline{\bm{\Omega}} +\underline{\bm{\gamma}} \right) {\bf M} + \underline{\bm{K}}^\top {\bf b^+}\,,\\
        {\bf b^-} &= \underline{\bm{\mathcal{C}}}{\bf b^+} + \underline{\bm{K}} {\bf M} \,,
    \end{split}
    \label{eq:CMT_main}
\end{equation}
where ${\bf M} = (A_1, X_\text{A$_1$,sc}, B_1, X_\text{B$_1$,sc}, X_\text{wc})^\top$ is a vector containing the amplitudes of the two photonic modes and the strongly and weakly coupled excitonic resonances, $X_\text{sc}$ and $X_\text{wc}$, respectively. The vectors $\bf b^\pm$ denote the amplitudes of the incoming ($+$) and outgoing ($-$) fields, while the matrix $\underline{\bm{\mathcal{C}}}$ represents all non-resonant scattering processes arising from Fabry--Pérot modes of an effective homogeneous slab, which accounts for the background dielectric responses of both the PhC and the TMD monolayer.

Each optical mode is characterized by its resonant energy $\hbar \omega_i$ and its total decay rate given by the sum of non-radiative ($\Gamma_i$) and radiative ($\gamma_i$) losses. Within this framework, the non-radiative contribution of the photonic modes is considered negligible ($\Gamma_{\text{A}_1} = \Gamma_{\text{B}_1} \approx 0$), due to the vanishing imaginary component of the dielectric function of the PhC slab. Furthermore, we neglect the direct radiative decay of the strongly coupled excitons. This is justified by Fig.~\ref{fig:SI_Photon_Coupled}c, which shows that it is dark excitons—located outside the lightcone—that couple to the A$_1$ and B$_1$ PhC modes. The two photonic modes are individually strongly coupled to their respective exciton resonance via the coupling strengths $g_{\text{A}_1/\text{B}_1}$. The matrix $\underline{\bm{\gamma}}$ contains the radiative decays $\gamma_i$ and the dissipative coupling $\tilde{g}= \sqrt{\gamma_{\text{A}_1/\text{B}_1} \gamma_X}$ \cite{suh2004temporal}. Under these assumptions, matrices $\hbar \underline{\bm{\Omega}}$ and $\underline{\bm{\gamma}}$ have the following form
\begin{equation}
    \hbar \underline{\bm{\Omega}} = \begin{pmatrix}
        \hbar \omega_{\text{A}_1} & g_{\text{A}_1} & 0 & 0 & 0 \\
        g_{\text{A}_1} & \hbar \omega_\text{X} - i\Gamma_\text{X} & 0 & 0 & 0 \\
        0 & 0 & \hbar \omega_{\text{B}_1} & g_{\text{B}_1} & 0 \\
        0 & 0 & g_{\text{B}_1} & \hbar \omega_\text{X} - i\Gamma_\text{X} & 0 \\
        0 & 0 & 0 & 0 & \hbar \omega_\text{X} - i\Gamma_\text{X} \label{eq:CMT_hbarOmega}
    \end{pmatrix},
\end{equation}
and 
\begin{equation}
    \underline{\bm{\gamma}} = i\begin{pmatrix}
        - \gamma_\text{A} & 0 & 0 & 0 & \tilde{g} \\
        0 & 0 & 0 & 0 &0 \\
        0 & 0 & - \gamma_\text{B} & 0 & 0\\
        0 & 0 & 0 & 0 & 0 \\
        \tilde{g} & 0 & 0 & 0 & -\gamma_\text{X}
    \end{pmatrix}.\label{eq:CMT_gamma}
\end{equation}
For a system satisfying mirror, reciprocity, and time-reversal symmetries \cite{fan2003temporal}, the coupling of the three radiative modes ($A_1$, $B_1$ and X$_\text{wc}$) to the incoming/outgoing ports $\bf b^\pm$ in the half-spaces above and below the PhC slab is described by the matrix:
\begin{equation}
    \underline{\bm{K}} = \begin{pmatrix}
        \sqrt{\gamma_\text{A}} & 0 & \sqrt{\gamma_\text{B}} & 0 & \sqrt{\gamma_\text{X}} \\
        \sqrt{\gamma_\text{A}} & 0 & \sqrt{\gamma_\text{B}} & 0 & \sqrt{\gamma_\text{X}}\label{eq:CMT_coupling}
    \end{pmatrix}.
\end{equation}

Transforming Eq.~\ref{eq:CMT_main} into the frequency domain yields the total scattering matrix of the structure \cite{fan2003temporal}:
\begin{equation}
    \underline{\bm{S}} = \begin{pmatrix} r && it \\ it && r\end{pmatrix} = \underline{\bm{C}} + \underline{\bm{K}} \left[i \left(\hbar\omega \underline{\bm{I}} - \hbar\underline{\bf{\Omega}} \right) + \underline{\bf{\gamma}} \right]^{-1} \underline{\bm{K}}^\top, 
    \label{eq:s}
\end{equation}
where $r$ and $t$ are the reflection and transmission coefficients, respectively, and $\underline{\bm{I}}$ is the identity matrix. Using Eq.~\ref{eq:s} alongside energy conservation, one can calculate the absorption of the system as 
\begin{equation}
A(\omega) = 1 - |r|^2 - |t|^2.\label{eq:abs}
\end{equation}
The exciton polariton energies can be obtained by solving $\det\underline{\bm{S}}=0$, leading to the following expression:
\begin{equation}
\begin{split}
    E_{{\text{A}_1/\text{B}_1}\pm} = \hbar \omega_{\text{P}_{\text{A}_1/\text{B}_1}\pm} &= \frac{\hbar\omega_\text{X} + \hbar\omega_{\text{A}_1/\text{B}_1}}{2} + i \frac{\gamma_\text{X,tot.} + \gamma_{\text{A}_1/\text{B}_1,\text{tot.}}}{2} \\
    &\pm \frac{1}{2}\underbrace{\sqrt{4g_{\text{A}_1/\text{B}_1}^2 + \left[\hbar \omega_\text{X} - \hbar\omega_{\text{A}_1/\text{B}_1} + i\left(\gamma_\text{X,tot.} - \gamma_{\text{A}_1/\text{B}_1,\text{tot.}}\right)\right]^2}}_{\hbar\Omega_{\text{R}}}, 
    \label{eq:polariton_energy} 
\end{split}
\end{equation}
where $\hbar\Omega_{\text{R}}$ is the Rabi splitting. The total decay rate, $\gamma_{i,\text{tot.}} = \gamma_i+\Gamma_i$, corresponds to half of the full width at half maximum of the individual uncoupled resonances.

To derive Eq.~2 in the main text, we neglect the contributions from the weakly coupled excitons in Eqs.~\ref{eq:CMT_hbarOmega} and \ref{eq:CMT_gamma}. Consequently, $\hbar \underline{\bm{\Omega}}$ and $\underline{\bm{\gamma}}$ are reduced to $4 \times 4$ matrices whose components remain the same with those defined in Eqs.~\ref{eq:CMT_hbarOmega} and \ref{eq:CMT_gamma} but neglecting the final row and column. Within this approximation, analogously to the derivation of Eqs.~\ref{eq:s} and \ref{eq:abs}, one finds the following polariton Elliott formula \cite{fitzgerald2022twist} for the absorption of the two upper and two lower polaritons under the assumption that upper and lower polariton branches are well separated:
\begin{equation}
A(\omega) = \sum_{i=\text{A,B},P=\text{UP,LP}}\frac{2 \gamma_{P,i} \Gamma_{P,i}}{\left(\gamma_{P,i} + \Gamma_{P,i}\right)^2 + (\hbar\omega - E_{i})^2},\label{eq:Elliot}
\end{equation}
where $\gamma_{P,i}$ and $\Gamma_{P,i}$ are the radiative and non-radiative decay rates of the polaritons determined by the expression $\Gamma_{P,i} = |C_\text{X}|^2 \Gamma_\text{X}$ and $\gamma_{P,i} = |C_\text{X}|^2 \gamma_\text{X}+ |C_{\text{Ph},i}|^2 \gamma_{\text{Ph},i}$. The Hopfield coefficients, $C_\text{X}$ and $C_{\text{Ph},i}$, are the components of the eigenvectors of the matrix $\hbar\omega \underline{\bm{I}} - \hbar\underline{\bm{\Omega}} + \underline{\bm{\gamma}}$. Note that Eq.~\ref{eq:Elliot} captures the 50\,\% absorption limit per mode.

\sisection{Perfect absorption under Gaussian beam illumination}

To calculate the absorption spectra shown in the main text, we assume that the incident illumination propagates normal to the surface of the PhC slab. However, typical reflection and transmission experiments are performed using microscope objectives, which focus the incident beam onto the sample and thus introduce a distribution of incidence angles rather than a single plane-wave excitation. To model these conditions and demonstrate that near-perfect absorption can be achieved under realistic illuminations, we consider an incident TE-polarized Gaussian beam, ${\bf E} = E\bf{\hat{y}}$, with an electric field amplitude
\begin{equation}
E(\theta) = \sqrt{2\pi w_0} \exp[-k_0^2\sin^2(\theta) \frac{w_0^2}{4}]. 
\end{equation}
Here, $w_0$ is the waist of the Gaussian beam, $k_0 = 2\pi/\lambda_0$ is the wavevector amplitude of the incident light and $\theta$ is the polar angle defined with respect to the $z$-axis.

\begin{figure}[t!]
    \centering
    \includegraphics[width=\linewidth]{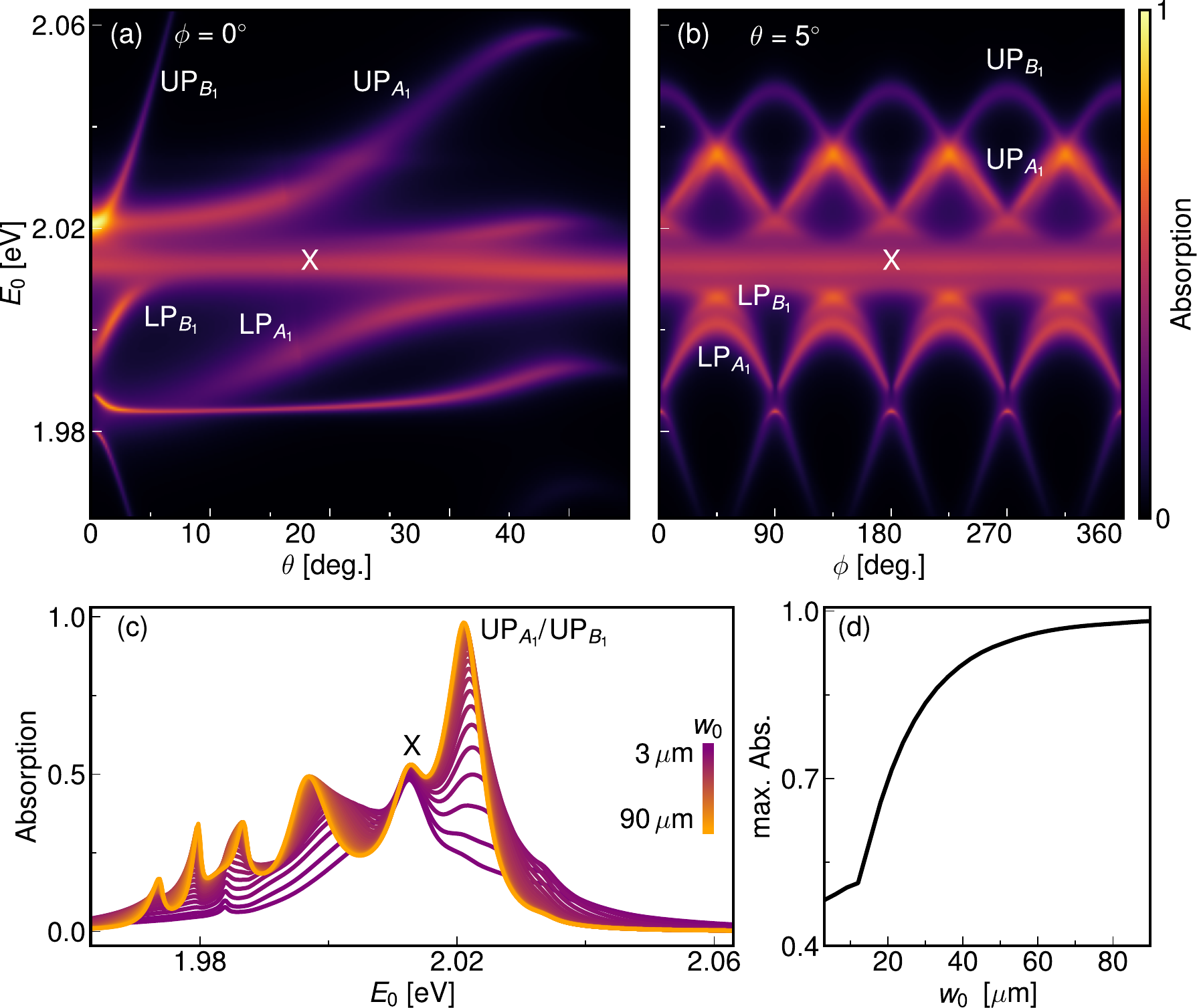}
    \caption{\textbf{(a)} Absorption spectra of the PhC/\ce{WS2} system versus incident energy and polar angle $\theta$. The dispersive branches of the two upper and lower polaritons as well as the weakly coupled exciton can be seen. Additional photonic modes that brighten for non-normal incidence are also visible. \textbf{(b)} Absorption spectra versus incident energy and the azimuthal angle $\phi$ for a fixed $\theta=5^{\circ}$ showing multiple dispersive polariton branches and a $90^{\circ}$-rotational symmetry, due to the square lattice of air holes. \textbf{(c)} Absorption spectra over the incident energy calculated for a Gaussian beam with beam waist $w_0$. Beam waist values between $3\,\mu$m (purple) and $90\,\mu$m (orange) are depicted as a color gradient. The larger the beam waist, the smaller the angular spread. \textbf{(d)} Maximum absorption as a function of the Gaussian beam waist. For $w_0 > 39$, $45$, and $75\,\mu$m, the maximum absorption exceeds $90$, $95$, and $97.5$\,\%, respectively.}
    \label{fig:SI_angle_dependence}
\end{figure}

We then calculate the absorption of the PhC--monolayer system using the same procedure as described in Ref.~\cite{kini2020suspended}. Specifically, we compute the TE- and TM-polarized reflection ($r_\text{TE}$, $r_\text{TM}$) and transmission ($t_\text{TE}$, $t_\text{TM}$) coefficients using RCWA for different polar and azimuthal ($\phi$) angles under plane-wave illumination. The resulting angle-dependent absorption spectra are shown in Figs.~\ref{fig:SI_angle_dependence}a and b. Using the angle-dependent reflection and transmission coefficients, we compute the reflectance ($R$) and transmittance ($T$) under Gaussian beam illumination as \cite{kini2020suspended}
\begin{align}
R &= \frac{\int_0^{\pi/2} \text{d}\theta \int_0^{2\pi} \text{d}\phi \,|E(\theta)|^2 \sin\theta \left(|r_\text{TE}|^2  \sin^2\phi +  |r_\text{TM}|^2 \cos^2\phi \right)}{\int_0^{\pi/2} \text{d}\theta \int_0^{2\pi} \text{d}\phi \,|E(\theta)|^2 \sin\theta},\label{eq:Angle_Dep_R}\\
T &= \frac{\int_0^{\pi/2} \text{d}\theta \int_0^{2\pi} \text{d}\phi \,|E(\theta)|^2 \sin\theta \left(|t_\text{TE}|^2  \sin^2\phi +  |t_\text{TM}|^2 \cos^2\phi \right)}{\int_0^{\pi/2} \text{d}\theta \int_0^{2\pi} \text{d}\phi\, |E(\theta)|^2 \sin\theta},\label{eq:Angle_Dep_T}
\end{align}
from which the total absorption can be calculated using energy conservation $(A = 1 - R - T)$.

Figure \ref{fig:SI_angle_dependence}c shows the simulated absorption spectrum of the system, calculated using Eqs.~\ref{eq:Angle_Dep_R} and \ref{eq:Angle_Dep_T}. We consider 30 beam waists ranging from $3$ to $90\,\mu$m in steps of $3\,\mu$m. The peak absorption of the energetically degenerate upper polaritons increase with the beam waist, reflecting the reduced angular spread of the broader Gaussian beam. This tendency is illustrated in Fig.~\ref{fig:SI_angle_dependence}a, which plots the maximum absorption as a function of beam waist. 

\newpage

\begin{figure}[b!]
    \centering
    \includegraphics[width=0.45\linewidth]{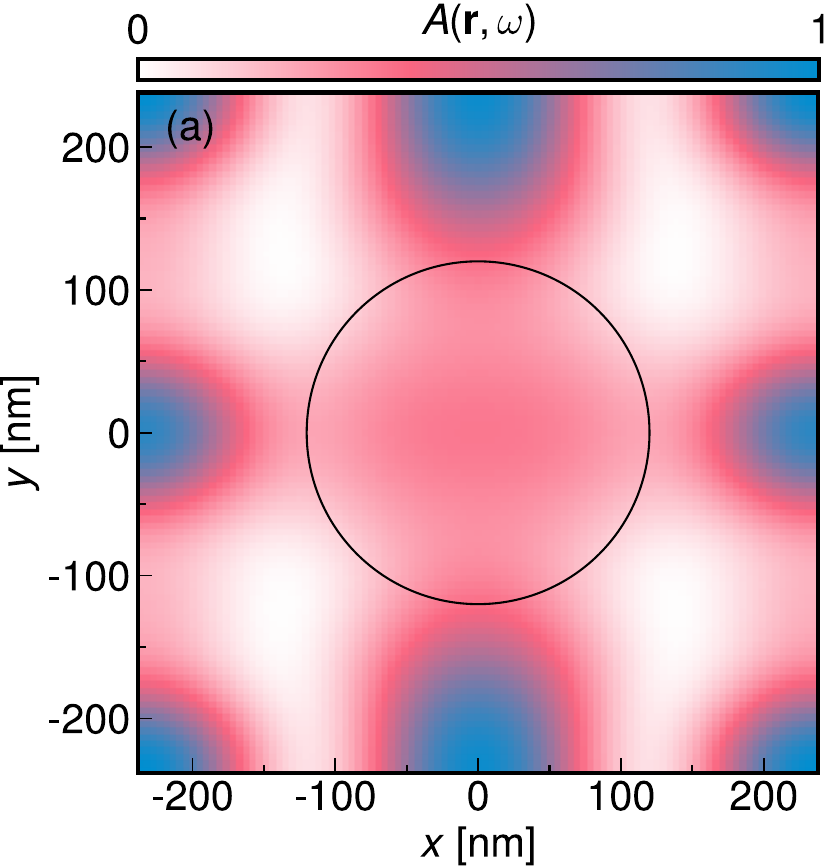}
    \caption{Normalized absorbed power density, $A(\bm{r},\omega)$, calculated via RCWA in the $xy$-plane of the \ce{WS2} monolayer, illustrating localized absorption within a single unit cell. The spatially resolved absorption is evaluated at the point of near-perfect absorption shown in Fig.~3a of the main text, where the two upper polaritons are degenerately critically coupled ($h=79.93\,$nm, $E_0=2.021$ eV, $f=0.80$, and $\Lambda = 478.1\,$nm). The black circle denotes the location of the PhC air hole.}
    \label{fig:SI_Poynting}
\end{figure}

\sisection{Spatial distribution of absorption}
To show the strong spatial confinement of absorption within the monolayer TMD in a single PhC unit cell, we evaluate the absorbed power density in the coupled PhC-monolayer system, defined as \cite{baffou2013thermo}:
\begin{equation}
A(\mathbf{r},\omega) = \frac{\omega}{2}\text{Im}[\varepsilon(\mathbf{r},\omega)]\varepsilon_0 |\mathbf{E}(\mathbf{r})|^2,
\end{equation}
where $\varepsilon_0$ is the permittivity of free space, $\varepsilon(\mathbf{r},\omega)$ is the spatially dependent dielectric function of the coupled system, and $\bf{E}(\bf{r})$ is the electric field obtained via RCWA calculations. Note that $A({\bf r}, \omega)=0$ outside the monolayer, since $\text{Im}[\varepsilon(\mathbf{r},\omega)] =0$ in the non-absorbing PhC. Consequently, absorption is strictly confined to the TMD, and the spatial variation in $A(\mathbf{r},\omega)$ directly reflects the highly spatially inhomogeneous electric field within the PhC unit cell. This can be seen in Fig.~\ref{fig:SI_Poynting}, which plots $A({\bf r}, \omega)$ in the $xy$-plane of the \ce{WS2} monolayer within a single PhC unit cell.  We can observe ``hot-spots'' of localized absorption on a 100-nm length scale, illustrating that PhCs hold potential for highly localized heating of 2D semiconductors.

\sisection{Temperature-dependent Linewidth of the \ce{WS2} 1s Exciton}
To model the temperature dependence of the \ce{WS2} 1s exciton linewidth, we interpolate the experimental data from Ref.~\citenum{meshulam2025temperature} using Eq.~4 of the main text. The fitted parameters are $\gamma_0 + \gamma = 5\,$meV, $c_1 =  12.5\,\mu$eVK$^{-1}$ and $c_2 = 14.6\,$meV, with a maximum deviation from the experimental data of $\approx9\,\%$.
\begin{figure}[hb!]
    \centering
    \includegraphics[width=0.6\linewidth]{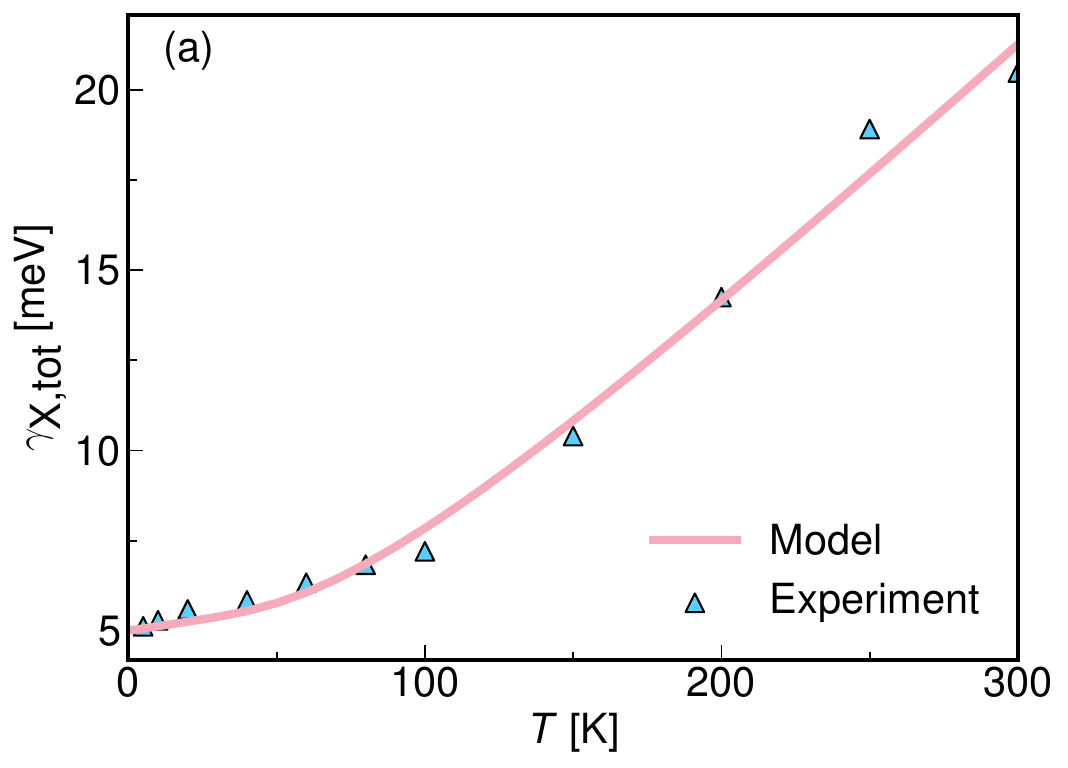}
    \caption{Temperature-dependent total linewidth of the \ce{WS2} A1s exciton calculated with Eq.~4 of the main text (solid line) using parameters fitted to experimental data from Ref.~\citenum{meshulam2025temperature} (blue triangles).}
    \label{fig:Exp_linewidth}
\end{figure}

\newpage

%\section{References for Supplementary Information}

\bibliography{bib}